\documentclass{aa}
\usepackage{graphicx}
\usepackage{amssymb}
\usepackage{natbib}
\usepackage{aalongtable}
\bibpunct{(}{)}{;}{a}{}{,}
\usepackage{lscape}
\usepackage{supertabular}
\usepackage{rotating}
\usepackage{setspace}
\usepackage{epsf}
\usepackage{txfonts}

\def\vsini{$V\!\sin i$}
\def\teff{$T_{\rm eff}$}

\def\logg{$\log~g$}

\def\omc{$\Omega/\Omega_{\rm{c}}$}

\def\top{$T_{\rm eff}^{\rm o}$}
\def\gop{$\log g_{\rm o}$}

\def\vsiniap{$V\!\sin i_{\rm~\!app.}$}
\def\vsinit{$V\!\sin i^{\rm true}$}
\def\kms{km~s$^{-1}$}
\def\tap{$T_{\rm eff}^{\rm app.}$}
\def\gap{$\log g_{\rm app.}$}

\def\rv{RV}
\def\ttms{$\frac{\tau}{\tau_{MS}}$}
\def\tms{$\tau_{MS}$}

\newcommand{\revised}{\bf}

\begin{document}
%\thesaurus{7}
%
\title{Effects of metallicity, star-formation conditions, and evolution in B and Be
stars.}
\subtitle{II: Small Magellanic Cloud, field of NGC\,330.}

\titlerunning{B and Be stars in the SMC}
\author{
C. Martayan \inst{1}
\and Y. Fr\'emat \inst{2}
\and A.-M. Hubert \inst{1}
\and M. Floquet \inst{1}
\and J. Zorec \inst{3}
\and C. Neiner \inst{1}
}
\offprints {C. Martayan}
\mail{christophe.martayan@obspm.fr}
\institute{
GEPI, UMR\,8111 du CNRS, Observatoire de Paris-Meudon, 92195 Meudon Cedex, France
\and Royal Observatory of Belgium, 3 avenue circulaire, 1180 Brussels, Belgium
\and Institut d'Astrophysique de Paris (IAP), 98bis boulevard Arago, 75014 Paris,
France
}
\date{Received /Accepted}

\abstract
{}
{ We search for effects of metallicity on B and Be stars in the Small and Large
Magellanic Clouds (SMC and LMC) and in the Milky Way (MW). We extend our
previous analysis of B and Be stars populations in the LMC to the SMC. The
rotational velocities of massive stars and the
evolutionary status of Be stars are examined with respect to their
environments.}
{ Spectroscopic observations of hot stars belonging to the young cluster
SMC-NGC\,330 and its surrounding region have been obtained with the VLT-GIRAFFE
facilities in MEDUSA mode. We determine fundamental parameters for B and Be
stars with the GIRFIT code, taking into account the effect of fast rotation, and
the age of observed clusters. We compare the mean \vsini~obtained by spectral
type- and mass-selection for field and cluster B and Be stars in the SMC with
the one in the LMC and MW.}
{ We find that (i) B and Be stars rotate faster in the SMC than in the LMC,
and in the LMC than in the MW; (ii) at a given metallicity, Be stars begin their
main sequence life with a higher initial rotational velocity than B stars.
Consequently, only a fraction of B stars that reach the ZAMS with a sufficiently
high initial rotational velocity can become Be stars; (iii) the distributions of
initial rotational velocities at the ZAMS for Be stars in the SMC, LMC and MW are
mass- and metallicity-dependent; (iv) the angular velocities of B and Be stars are
higher in the SMC than in the LMC and MW; (v) in the SMC and LMC, massive Be stars
appear in the second part of the main sequence, contrary to massive Be stars in the
MW.}
{}


\maketitle

\section{Introduction}

The origin of the Be phenomenon has given rise to long debates. 
Whether it is linked to stellar evolution or initial formation 
conditions remains a major issue. Thus, finding out  differences in
the physical properties of B and Be stars populations belonging to
environments with different metallicity could provide new clues to understand
the Be phenomenon.

To investigate the influence of metallicity,  star-formation
conditions and stellar evolution on the Be phenomenon, we have undertaken an
exhaustive study of B and Be stars belonging to young clusters or field of the
Small and Large Magellanic Clouds (SMC and LMC), because these galaxies have a
lower metallicity  than the Milky Way (MW). For this purpose we made use of the
new  FLAMES-GIRAFFE instrumentation installed at the VLT-UT2 at ESO,
which is particularly well suited, in MEDUSA mode, to obtain high quality
spectra of large samples needed for the study of stellar populations. In
Martayan et al. (2006a, hereafter  M06), we reported on the
identification of 177 B and Be stars  belonging to the young cluster LMC-NGC
2004 and its surrounding region. In Martayan et al. (2006b, hereafter Paper
 I),  we determined  fundamental parameters of a large
fraction of the sample in the  LMC, taking into account rotational
effects (stellar flattening, gravitational darkening) when appropriate.
 We then investigated the effects of metallicity on rotational
velocities. We concluded that Be stars begin their life on the main
sequence (MS) with a  higher initial velocity than B stars. 
Moreover, this initial velocity is sensitive to the metallicity. Consequently,
only a fraction of the B stars that reach the ZAMS with a sufficiently high initial
rotational velocity can become Be stars. However, no
clear influence of metallicity on the rotational velocity of B stars was found. 

The present paper deals with a large sample of B and Be stars in the SMC, which
has a lower metallicity than the LMC. With the determination of
fundamental parameters, and the study of the evolutionary status and abundances,
 we aim at confirming and enlarging our results derived from the study of B
and Be stars in Martayan et al. (2005a) and from the LMC (Paper I).

\section{Observations}

This work makes use of spectra obtained with the multifibre  VLT-FLAMES/GIRAFFE
spectrograph in Medusa mode (131 fibres) at medium resolution (R=6400) in setup
LR02 (396.4 - 456.7 nm). Observations  (ESO runs 72.D-0245A and 72.D-0245C) were
carried out in the young cluster SMC-NGC\,330 and in its surrounding field, as
part of the Guaranteed Time Observation programmes  of the Paris Observatory
(P.I.: F. Hammer).  The observed fields (25\arcmin~in diameter) are centered at
$\alpha$(2000) = 00h 55mn 15s, $\delta$(2000) = -72$^{\circ}$ 20\arcmin~
00\arcsec~  and $\alpha$(2000) = 00h 55mn 25s,   $\delta$(2000) = -72$^{\circ}$
23\arcmin~ 30\arcsec. Besides the young cluster NGC\,330, this field contains
several high-density groups of stars (NGC306, NGC299, OGLE-SMC99, OGLE-SMC109,
H86 145, H86 170, Association [BS95]78, Association SMC ASS39).  Note that we
corrected the coordinates of NGC299 given in Simbad (CDS) with EIS
coordinates (Momany et al. 2001).  Spectra have been obtained on
October 21, 22, and 23, 2003 and September 9 and 10, 2004; at those dates, the
heliocentric velocity is respectively 7 and 12 \kms. The strategy and
conditions of observations, as well as the spectra reduction 
procedure, are described in M06. A sample of 346 stars
has been observed within the two observing runs. Since the V magnitude of the
selected targets ranges from 13.5 to 18.8 mag, the integration time varied 
between 1h and 2h. However, as the seeing was not optimal during the first run, 
the S/N ratio is only about 50 on average, with individual values ranging 
from 20 to 130.

We pre-selected 11544 B-type star candidates with 14$\leq$V$\leq$18
and a colour index B-V$<$0.35, among the 192437 stars listed in the EIS SMC5
field by the EIS team (Momany et al. 2001), keeping in mind the intrinsic value
E(B-V)=0.08 (Keller et al. 1999) for the SMC. We then observed three fields with
VLT-FLAMES/GIRAFFE. Since for each field maximum 130 stars can be observed, we 
collected data for 346 objects among the 5470 B-type star candidates located in
these fields for the selected magnitude range. The ratio of observed to observable
B-type stars in the GIRAFFE fields is thus 6.3\%. This represents a
statistically significant sample. From the observations we confirm the B
spectral type for 333 of the 346 stars. Of the remaining 13 objects, 4 are O
stars, 6 are A stars, 1 is a cold supergiant, 1 is a planetary nebula, and 1 is a HB[e]. 
The 333 B-type objects further include
131 Be stars.

\begin{figure}[!htbp]
\centering
\resizebox{\hsize}{!}{\includegraphics[angle=-90]{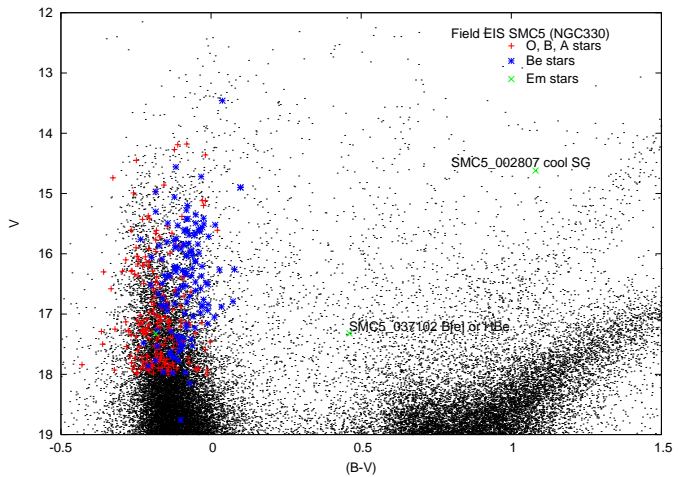}}
\caption{V versus (B-V) colour diagram from EIS
photometry in the EIS SMC\,5 field. The '.' symbols correspond to all stars in
this field. '*' show the Be stars, '+' the O-B-A stars and 'x' the other 
emission-line stars in the sample.}
\label{figcoul}
\end{figure}

The V versus B-V colour diagram (Fig.~\ref{figcoul}), derived from EIS
photometry (Momany et al. 2001), shows the O, B, A and Be stars in our sample 
compared to all the stars in the EIS-SMC\,5 field. 
%The locations of the
%{\revised targets} are shown in the SMC\,5 field from the EIS pre-FLAMES survey
%(Fig.~\ref{figure0}).

%
\section{Determination of fundamental parameters}
\label{FPD}

As in Paper I we make use of the GIRFIT least-square procedure
(Fr\'emat et al. 2006) to derive the fundamental parameters: effective
temperature (\teff), surface gravity (\logg), projected rotational velocity
(\vsini) and radial velocity (RV). This procedure fits the observations with
theoretical spectra interpolated in a grid of stellar fluxes computed with the
SYNSPEC programme and from model atmospheres calculated with TLUSTY (Hubeny \&
Lanz 1995, see references therein) or/and with ATLAS9 (Kurucz 1993; Castelli et
al. 1997). The grid of model atmospheres we use to build the GIRFIT input of
stellar fluxes  is obtained in the same way as in our LMC study, but for the
metallicity of the SMC. 

The metallicities of the model atmospheres are chosen to be as close as possible to the
NGC\,330 average value, $[m/H]~=~-0.6$ (where $[m/H] = \log(m/H)_{\rm SMC} -
\log(m/H)_\odot$), estimated from Jasniewicz \& Th\'evenin (1994). The
Kurucz and OSTAR 2002 models we use are therefore those calculated with a [m/H] close to -0.6.
Finally, the complete input flux grid is built assuming the averaged element abundances
derived by Jasniewicz \& Th\'evenin (1994) for C, Mg, Ca, Ti, Cr, Mn and Fe. The other elements, except
hydrogen and helium, are assumed to be underabundant by $-0.6$ dex relative to the
Sun.

It is worth noting that GIRFIT does not include the effects of fast rotation. 
Therefore, for rapidly rotating stars, we need to correct the stellar
parameters with the FASTROT computer code  (Fr\'emat et al. 2005) assuming a solid-body-type rotation. 
We then obtain the 'parent non-rotating counterpart' (pnrc; see Fr\'emat et al.
2005) stellar parameters (\top, \gop, \vsinit) for a given \omc.

For a more detailed description of the grid of model atmospheres we
use, the fitting criteria we adopt in the GIRFIT procedure, and the
correction for fast rotation we apply on fundamental parameters of Be stars, we
refer the reader to Paper I (Sect. 3).

Finally, we determine the spectral classification of each star with
two methods. The calibration we establish to estimate these spectral types is
described in Paper I. The agreement between the two methods is not as good as
for the stars observed in the LMC (Paper I), because the observations for
the SMC have a lower S/N ratio.

\section{ Stellar parameters of the sample stars}

\begin{table*}[!tbph]
\centering
\footnotesize{
\caption{\label{BPNM} Fundamental parameters for O, B, A stars in the SMC.
The name, coordinates ($\alpha$(2000), $\delta$(2000)), V magnitude and (B-V) colour index of stars are
taken from the EIS catalogues. The effective temperature \teff~is given in K, \logg~in dex, \vsini~in \kms~and the \rv~in \kms.
For each parameter the 1$\sigma$ error is given.
The abbreviation ``CFP'' is the spectral type and luminosity
classification determined from fundamental parameters (method 2), whereas
'CEW' is the spectral type and luminosity classification determined from EW
diagrams (method 1). The localization in clusters is indicated in the last column: 
cl0 for NGC\,330 (0h 56m 19s -72$^{\circ}$ 27\arcmin~52\arcsec),
cl1 for H86 170 (0h 56m 21s -72$^{\circ}$ 21\arcmin~12\arcsec),
cl2 for [BS95]78 (0h 56m 04s -72$^{\circ}$ 20\arcmin~12\arcsec),
cl3 for the association SMC ASS 39 (0h 56m 6s -72$^{\circ}$ 18\arcmin~00\arcsec),
cl4 for OGLE-SMC109 (0h 57m 29.8s -72$^{\circ}$ 15\arcmin~51.9\arcsec),
cl5 for NGC299 (0h 53m 24.5s -72$^{\circ}$ 11\arcmin~49\arcsec) corrected coordinates,
cl6 for NGC306 (0h 54m 15s -72$^{\circ}$ 14\arcmin~30\arcsec),
cl7 for H86 145 (0h 53m 37s -72$^{\circ}$ 21\arcmin~00\arcsec),
cl8 for OGLE-SMC99 (0h 54m 48.24s -72$^{\circ}$ 27\arcmin~57.8\arcsec).
The full table is available online.}
\begin{tabular}{@{\ }l@{\ \ \ }l@{\ \ \ }l@{\ \ \ }l@{\ \ \ }l@{\ \ \ }l@{\ \ \ }l@{\ \ \ }l@{\ \ \ }l@{\ \ \ }l@{\ \ \ }l@{\ \ \ }l@{\ \ \ }l@{\ }}
\hline
\hline
Star & $\alpha$(2000) & $\delta$(2000) & V & B-V & S/N & \teff  & \logg  & \vsini & RV & CFP & CEW & com.\\
\hline
SMC5\_000351 & 00 53 32.810 & -72 26 43.70 & 17.75 & -0.15 & 20 & 16500$\pm$1600 & 4.4$\pm$0.2 & 312$\pm$47 & 158 $\pm$10 & B3V & B2.5IV &  \\
SMC5\_000398 & 00 54 02.700 & -72 25 40.60 & 14.27 & -0.12 & 90 & 13500$\pm$400 & 2.7$\pm$0.1 & 104$\pm$10 & 138 $\pm$10 & B5II-III & B0.5III &  \\
SMC5\_000432 & 00 53 19.101 & -72 24 48.76 & 16.87 & -0.14 & 60 & 15500$\pm$800 & 3.7$\pm$0.2 & 197$\pm$16 & 156 $\pm$10 & B3IV & B2III &  \\
... & ... & ... & ...\\
\hline
\end{tabular}
}
\end{table*}

 In this section we present the stellar parameters and spectral
classification we obtained for O-B-A and Be stars.

\subsection{ Fundamental parameters of O-B-A stars}

\label{subsec:HRtracks}

Early-type stars that do not show intrinsic emission lines in their spectrum and
have not been detected as spectroscopic binaries are listed in Table~\ref{BPNM},
sorted by their EIS catalogue number. The fundamental parameters \teff, \logg,
\vsini, and \rv~obtained by fitting the observed spectra, as well as the
spectral classification deduced on one hand from \teff--\logg~plane calibration
(CFP determination, see Paper I) and on the other hand from
equivalent width diagrams (CEW determination, see Paper I), are
reported in Table~\ref{BPNM}. The heliocentric velocities (7 and 12 \kms)
have been subtracted from the radial velocities.

To derive the luminosity, mass and radius of O, B, and A stars from
their fundamental parameters, we interpolate in the HR-diagram grids (Schaller
et al. 1992) calculated for the SMC metallicity (Z = 0.001; Maeder et al. 1999
and references therein) and for stars without rotation.

We estimate  the mean radius, mean mass and mean \vsini~ in various
mass bins (e.g. 5 $<$ M $<$ 7 M$\odot$, 7 $<$ M $<$ 9 M$\odot$, etc). We then
obtain a mean equatorial velocity for a random angle distribution using formulae
published in Chauville et al. (2001) and Paper I). 

For B stars, $<$\vsini$>$ is close to 160 km~s$^{-1}$, thus
$V_{e}/V_{c}$ $\simeq$ 43\% and \omc $\simeq$ 58\%. As the effects of fast
rotation on the spectra are only significant for \omc $>$ 60\% (Fr\'emat et al.
2005), we do not need to correct the fundamental parameters of B
stars for fast rotation effects. This justifies the use of non-rotating 
models. Although some B stars do have a high
\vsini~($>$350 \kms), the accuracy on the parameters determination is generally
low for these stars, and thus we decide not to introduce corrections.
Since the value of the averaged \omc~is at the limit 
at which the spectroscopic effects of fast rotation appear,
we however expect that a significant part of the B stars in the
sample will be apparently more evolved due to gravitational darkening.

The obtained luminosity, mass, radius, and age of most O, B, and A stars of
the sample are given in Table~\ref{MLRBPNM}. The position of these stars in the
HR diagram is shown in Fig.~\ref{hrPNM}.

\begin{table}[!tbph]
\centering
\footnotesize{
\caption{\label{MLRBPNM} Parameters $\log(L/L_{\odot}$), $M/M_{\odot}$ et $R/R_{\odot}$
interpolated or calculated for our sample of O, B, A stars in the SMC 
from HR diagrams published in Schaller et al. (1992) for Z=0.001.
The full table is available online.}
\begin{tabular}{@{\ }l@{\ \ \ }l@{\ \ \ }l@{\ \ \ }l@{\ \ \ }l@{\ \ \ }l@{\ \ \ }l@{\ \ \ }l@{\ \ \ }l@{\ \ \ }l@{\ \ \ }l@{\ \ \ }l@{\ \ \ }l@{\ }}
\hline
\hline
Star & $\log(L/L_{\odot})$ & $M/M_{\odot}$ & $R/R_{\odot}$ & age (Myears) \\
\hline
SMC5\_000351 & 2.4 $\pm$ 0.4 & 3.4 $\pm$ 0.5 & 1.9 $\pm$ 0.3 & 125 $\pm$6\\
SMC5\_000398 & 4.2 $\pm$ 0.4 & 9.7 $\pm$ 0.5 & 24.8 $\pm$ 2.5 & 26 $\pm$3\\
SMC5\_000432 & 3.1 $\pm$ 0.4 & 4.5 $\pm$ 0.5 & 5.0 $\pm$ 1.0 & 119 $\pm$6\\
... & ... & ... & ...\\
\hline
\end{tabular}
}
\end{table}

\begin{table*}[tbph!]
\footnotesize{
\caption{See the caption of Table~\ref{BPNM}. In addition, MHF[SX]XXXXX is given for stars in our catalogue
of emission line stars, and in the last column, the letter ``k'' followed by a number corresponds to 
the star number in Keller et al. (1999). Moreover the ``\#'' indicates that the star 
was pre-selected in our catalogue of emission line stars (unpublished).
In the last column the spectral classification from the SIMBAD database is also given.
The last three lines correspond to emission-line objects which are not Be stars.
The full table is available online.}
\label{tableBePNM}
\centering
\begin{tabular}{@{\ }l@{\ \ \ }l@{\ \ \ }l@{\ \ \ }l@{\ \ \ }l@{\ \ \ }l@{\ \ \ }l@{\ \ \ }l@{\ \ \ }l@{\ \ \ }l@{\ \ \ }l@{\ \ \ }l@{\ \ \ }l@{\ }}
\hline
\hline
Star & $\alpha$ (2000) & $\delta$ (2000) & V & B-V & S/N & \tap  & \gap  & \vsiniap & RV & CFP & comm.\\
\hline
MHF[S61]47315 & 0 54 49.559 & -72 24 22.35 & \_ & \_ & 120 & 30000$\pm$800 & 3.4$\pm$0.1 & 370$\pm$10 & 160$\pm$10 & B0IV & \# \\
MHF[S61]51066 & 0 54 50.936 & -72 22 34.63 & \_ & \_ & 130 & 22500$\pm$600 & 3.3$\pm$0.1 & 415$\pm$10 & 130$\pm$10 & B1III & \# \\
SMC5\_000476 & 0 53 23.700 & -72 23 43.80 & 16.36 & -0.15 & 40 & 18500$\pm$1400 & 4.0$\pm$0.2 & 309$\pm$30 & 110$\pm$10 & B2V &  \\
... & ... & ... & ...\\
\hline
\end{tabular}
}
\end{table*}

\subsection{Fundamental parameters of Be stars}
\subsubsection{ Apparent fundamental parameters}

The sample (131 Be stars) includes 41 known
Be stars from Keller et al. (1999) and from Grebel et al. (1992), for which the
H$\alpha$ emissive character has been confirmed in this work, and 90
new Be stars. Three H$\alpha$ emission line stars mentioned in Keller et al.
(1999) are not Be stars: the star SMC5\_2807 or KWBBe044 is a cool supergiant
and a binary, the star SMC5\_37102 or KWBBe485 is a possible HB[e], and the star
SMC5\_81994 or KWBBe4154 is a planetary nebula. 

The apparent fundamental parameters (\tap, \gap, \vsiniap, and \rv) we derive
for these stars are reported in Table~\ref{tableBePNM}. The spectral
classification derived from apparent fundamental parameters is also given in the
last column of the Table. Without correction for fast rotation nearly all Be
stars seem to be sub-giants or giants.

The apparent luminosity, mass, radius, and age of Be stars are
derived in the same way as for O, B and A stars (see
Sect.~\ref{subsec:HRtracks}), from their apparent fundamental
parameters. The apparent position of Be stars in the H-R diagram is shown in
Fig.~\ref{hrPNM} and the  corresponding luminosities, masses, and
radii are given in Table~\ref{LMRBePNMapp}.

\begin{table}[!tbph]
\centering
\footnotesize{
\caption{\label{LMRBePNMapp} Apparent parameters $\log(L/L_{\odot}$), $M/M_{\odot}$, and $R/R_{\odot}$ 
interpolated or calculated for Be stars in the SMC from HR diagrams published in Schaller et al. (1992) for Z=0.001.
The full table is available online.}
\begin{tabular}{@{\ }l@{\ \ \ }l@{\ \ \ }ll@{\ \ \ }lll@{\ \ \ }ll@{\ \ \ }l@{\ \ \ }l@{\ \ \ }l@{\ \ \ }l@{\ }}
\hline
\hline
Star & $\log(L/L_{\odot})$ & $M/M_{\odot}$ & $R/R_{\odot}$ & age Myears\\
MHF[S9]47315 & 5.3 $\pm$ 0.4 & 23.9 $\pm$ 1.5 & 16.0 $\pm$ 1.5 & 7.6 $\pm$ 1\\
MHF[S9]51066 & 4.6 $\pm$ 0.4 & 13.1 $\pm$ 1.0 & 14.1 $\pm$ 1.5 & 16.2 $\pm$ 3\\
SMC5\_000476 & 3.2 $\pm$ 0.4 & 5.2 $\pm$ 0.5 & 3.8 $\pm$ 0.5 & 77.2 $\pm$ 6\\
... & ... & ... & ...\\
\hline
\end{tabular}
}
\end{table}

\subsubsection{Fundamental parameters corrected for rapid rotation}
\label{befastcalculs}

The pnrc fundamental parameters (\top, \gop, \vsinit) of Be stars we obtain after
correction with FASTROT are given in Table~\ref{tableBePNM95} for different
rotation rates \omc. We estimate the rotation rate \omc~to be used for the
selection of the most suitable pnrc fundamental parameters of Be stars in the
SMC as in Paper I. We obtain $V_{e}/V_{c}$~$\simeq$~87\% and
\omc~$\simeq$~95\% on average.

\begin{table*}[tbph!]
\footnotesize{
\caption{Fundamental parameters for Be stars in the SMC corrected from the effects of
fast rotation assuming different rotation rates (\omc).
The most suitable corrections are those corresponding to \omc=95\%. 
The units are K for \top, dex for \gop, and \kms~for \vsinit.
The full table is available online.}
\label{tableBePNM95}
\centering
\begin{tabular}{@{\ }l@{\ \ \ }l@{\ \ \ }l@{\ \ \ }l@{\ \ \ }l@{\ \ \ }l@{\ \ \ }l@{\ \ \ }l@{\ \ \ }l@{\ \ \ }l@{\ \ \ }l@{\ \ \ }l@{\ \ \ }l@{\ }}
\hline
\hline
star &     & \omc=85\% &       &\vline&     & \omc=90\%&       & \vline&     &\omc=95\% & \\      
SMC &  \top  & \gop     & \vsinit  &\vline &  \top   &  \gop   & \vsinit  &\vline &  \top   & \gop &  \vsinit  \\ 
\hline
MHF[S9]47315 & 33000 $\pm$800 & 3.7 $\pm$0.1 & 383 $\pm$10 & \vline & 34000 $\pm$800 & 3.7 $\pm$0.1 & 390 $\pm$10 & \vline & 32500 $\pm$800 & 3.8 $\pm$0.1 & 396 $\pm$10 \\
MHF[S9]51066 & 26500 $\pm$600 & 3.7 $\pm$0.1 & 428 $\pm$10 & \vline & 27500 $\pm$600 & 3.8 $\pm$0.1 & 437 $\pm$10 & \vline & 26000 $\pm$600 & 3.8 $\pm$0.1 & 450 $\pm$10 \\
SMC5\_000476 & 20000 $\pm$1400 & 4.3 $\pm$0.2 & 321 $\pm$30 & \vline & 20000 $\pm$1400 & 4.3 $\pm$0.2 & 328 $\pm$30 & \vline & 20000 $\pm$1400 & 4.3 $\pm$0.2 & 336 $\pm$30 \\
... & ... & ... & ...\\
\hline
\end{tabular}
}
\end{table*}

\begin{figure}[!ht]
\centering
\resizebox{\hsize}{!}{\includegraphics[angle=-90]{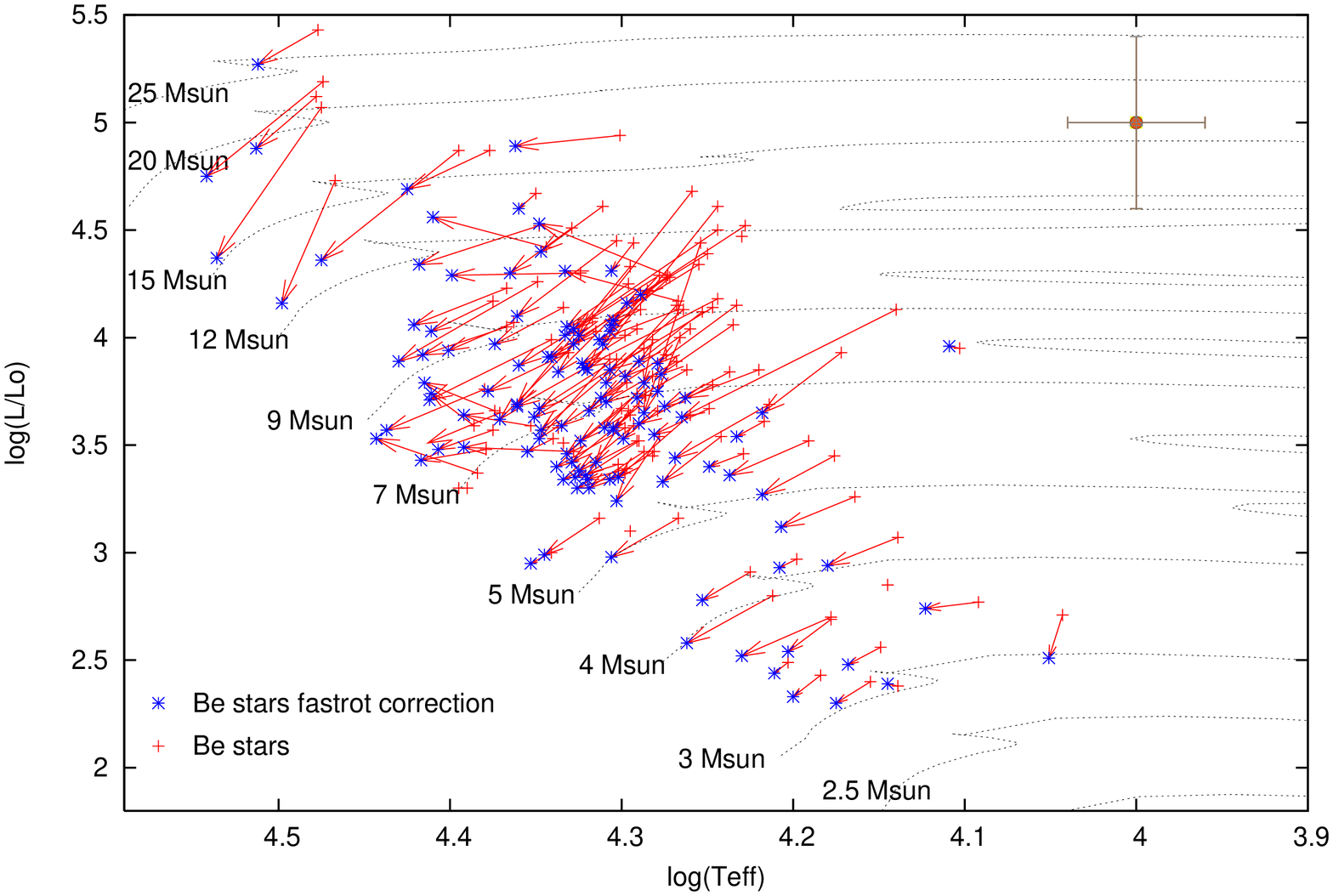}}\\
\resizebox{\hsize}{!}{\includegraphics[angle=-90]{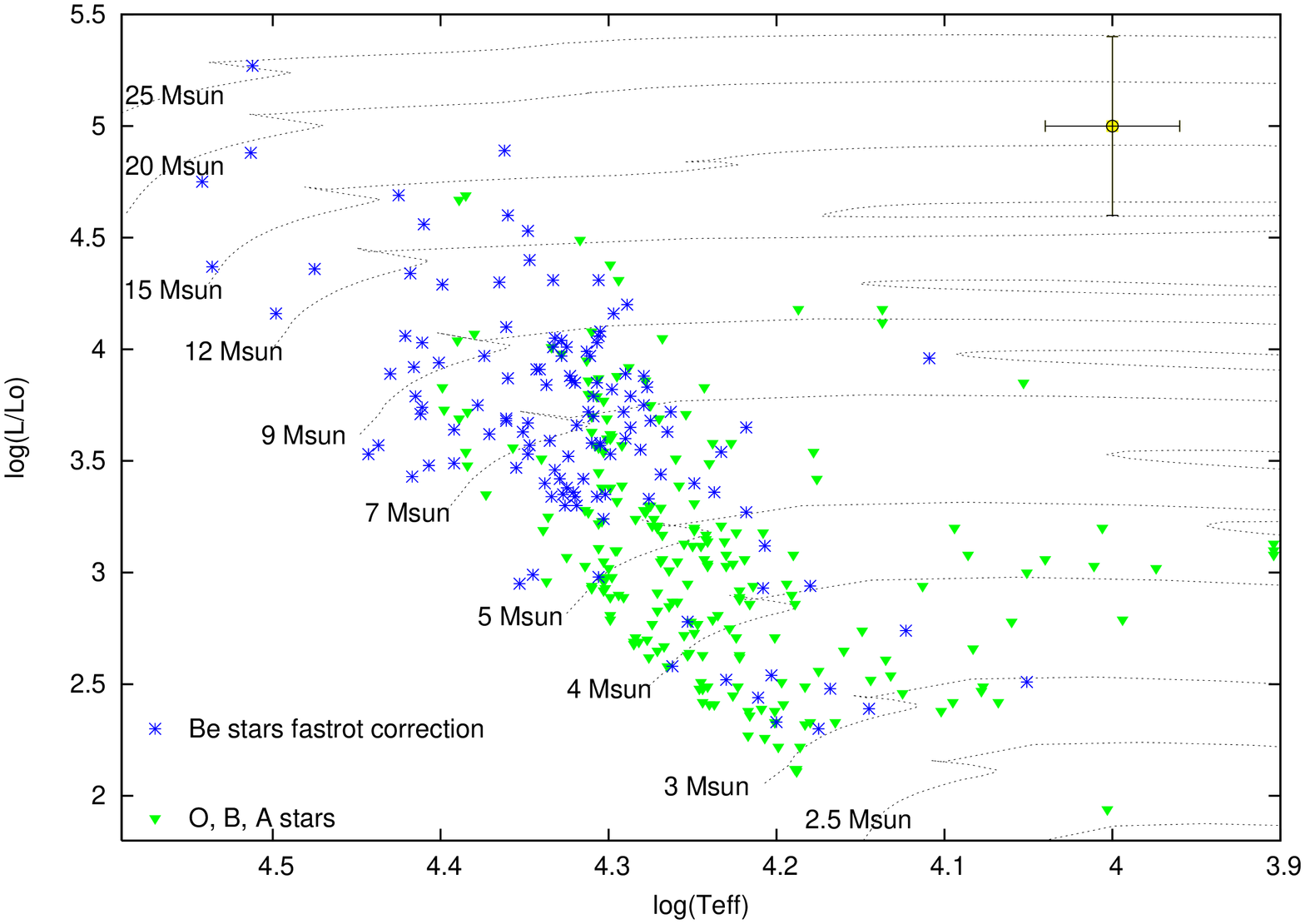}}
\caption{HR diagrams for the studied B and Be stars.  
Top: The effects of fast rotation are taken into account with \omc~=
95\%  for Be stars.
Bottom: B stars and fast rotators (Be stars) corrected for their fast rotation.
Common: The adopted metallicity for the SMC is Z=0.001.  
Red '+' represent Be stars with their apparent parameters, blue '*' Be stars 
corrected with FASTROT with \omc~=95\%, and green triangles B stars.  
Typical error bars are shown in the upper right corner of the figure.
Evolutionary tracks come from Schaller et al. (1992).}
\label{hrPNM}
\end{figure}

As previously, but with the pnrc fundamental parameters corresponding
to the rotation rate \omc~=~95\%, we derive the luminosity
$\log(L/L_{\odot})$, mass $M/M_{\odot}$, and radius $R/R_{\odot}$ for Be stars.
These parameters are given in Table~\ref{LMRBePNM95}. After correction for rapid
rotation, Be stars globally shift in the HR diagram towards lower luminosity and
higher temperature, as illustrated in Fig.\ref{hrPNM}. It clearly demonstrates
that Be stars are less evolved than their apparent fundamental parameters could
indicate.

\begin{table}[tbph!]
\centering
\footnotesize{
\caption{\label{LMRBePNM95} 
Parameters $\log(L/L_{\odot}$), $M/M_{\odot}$, $R/R_{\odot}$ and age of Be stars in the SMC, obtained by interpolation in the evolutionary
tracks published in Schaller et al. (1992) with Z=0.001 with fundamental parameters corrected for fast rotation effects with \omc=95\%.
The full table is available online.}
\begin{tabular}{@{\ }l@{\ \ \ }l@{\ \ \ }l@{\ \ \ }l@{\ \ \ }l@{\ \ \ }l@{\ \ \ }l@{\ \ \ }l@{\ \ \ }l@{\ \ \ }l@{\ \ \ }l@{\ \ \ }l@{\ \ \ }l@{\ }}
\hline
\hline
Star & $\log(L/L_{\odot})$ & $M/M_{\odot}$ & $R/R_{\odot}$ & age Myears\\
\hline
MHF[S9]47315 & 5.0 $\pm$ 0.4 & 21.3 $\pm$ 1.5 & 10.2 $\pm$ 1.5 & 7.9 $\pm$1\\
MHF[S9]51066 & 4.4 $\pm$ 0.4 & 12.0 $\pm$ 1.0 & 7.6 $\pm$ 1.5 & 17.49 $\pm$3\\
SMC5\_000476 & 3.0 $\pm$ 0.4 & 5.0 $\pm$ 0.5 & 2.5 $\pm$ 0.5 & 51.42 $\pm$6\\
... & ... & ... & ...\\
\hline
\end{tabular}
}
\end{table}

\subsubsection{Spectral lines saturation}

According to Townsend et al. (2004) and Fr\'emat et al. (2005, Figs. 5 and 6),
there may be a saturation effect of the FWHM of spectral lines
for the highest angular velocities (\omc)  that hampers the estimate of \vsini.
However, the magnitude of this effect strongly depends on stellar 
parameters and on the studied line-transitions. Multiple line fitting,
as we perform in our study, allows therefore to reduce the impact of
the saturation (due to the gravitational darkening) and
to correct it with FASTROT. This is confirmed by the fact that, for the SMC,
we report apparent $\frac{V_{e}}{V_{c}}$ ratios which are significantly above the
expected limit where saturation should appear (i.e. $\sim$0.80).

\subsection{Characteristics of the sample}
\label{charsample}

To characterize the sample of stars, we study the distribution in spectral
types, luminosity classes, and masses for stars in clusters and in the 
field.
Note that the method with equivalent widths (CEW) fails to give a reliable 
spectral classification for the few hotter (late O) and cooler (B5-A0) stars in
the sample. Moreover, for Be stars, the spectral classification determination is
only performed using the derived apparent fundamental parameters (CFP), since the 
emission contamination, often present in H$\gamma$ and in
several cases in the \ion{He}{i} 4471 line, makes the first method particularly
inappropriate for early Be stars.

\subsubsection{Distributions in spectral types and luminosity classes}

\begin{figure}[!ht]
\centering
\resizebox{\hsize}{!}{\includegraphics[angle=-90]{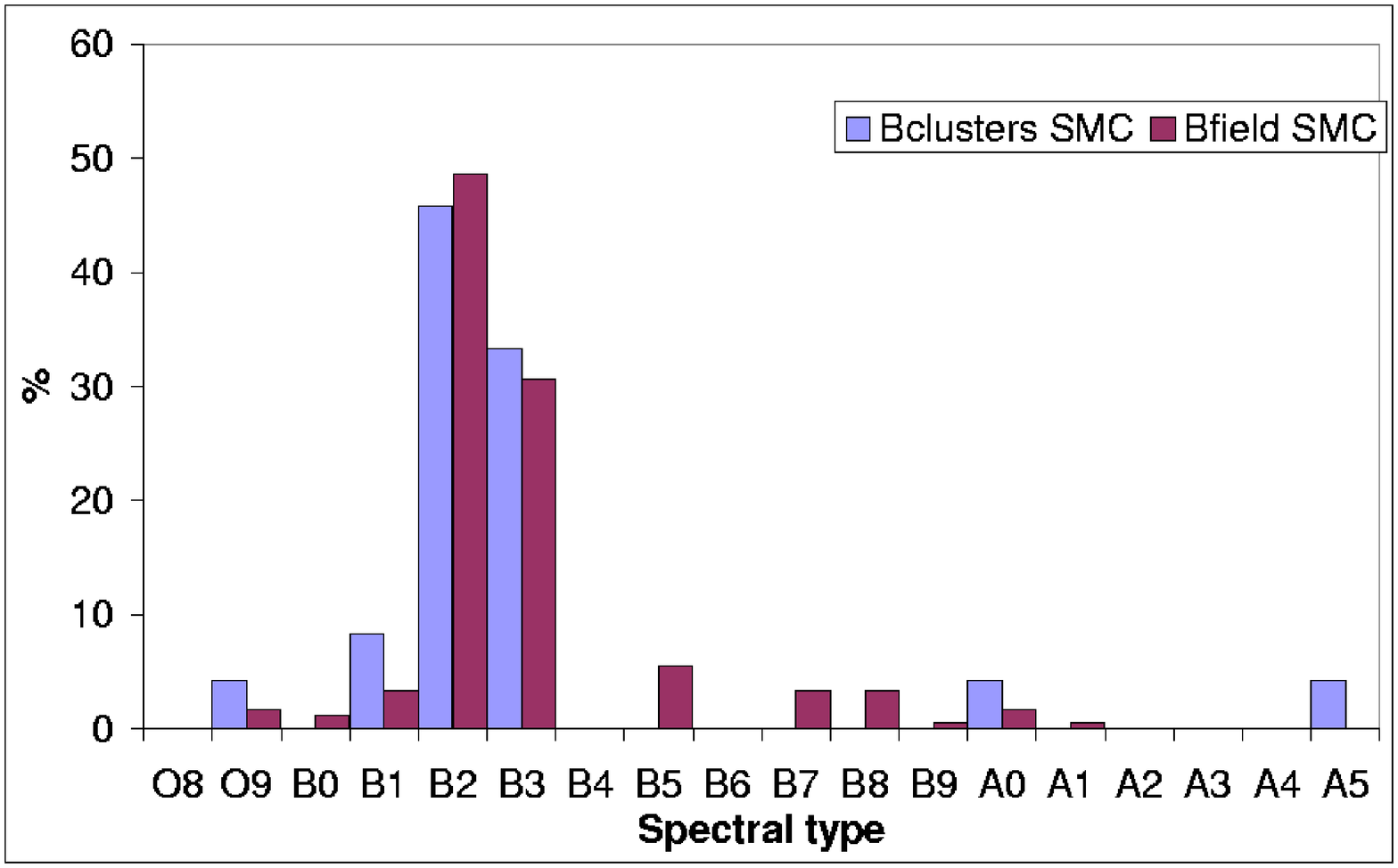}}\\
\resizebox{\hsize}{!}{\includegraphics[angle=-90]{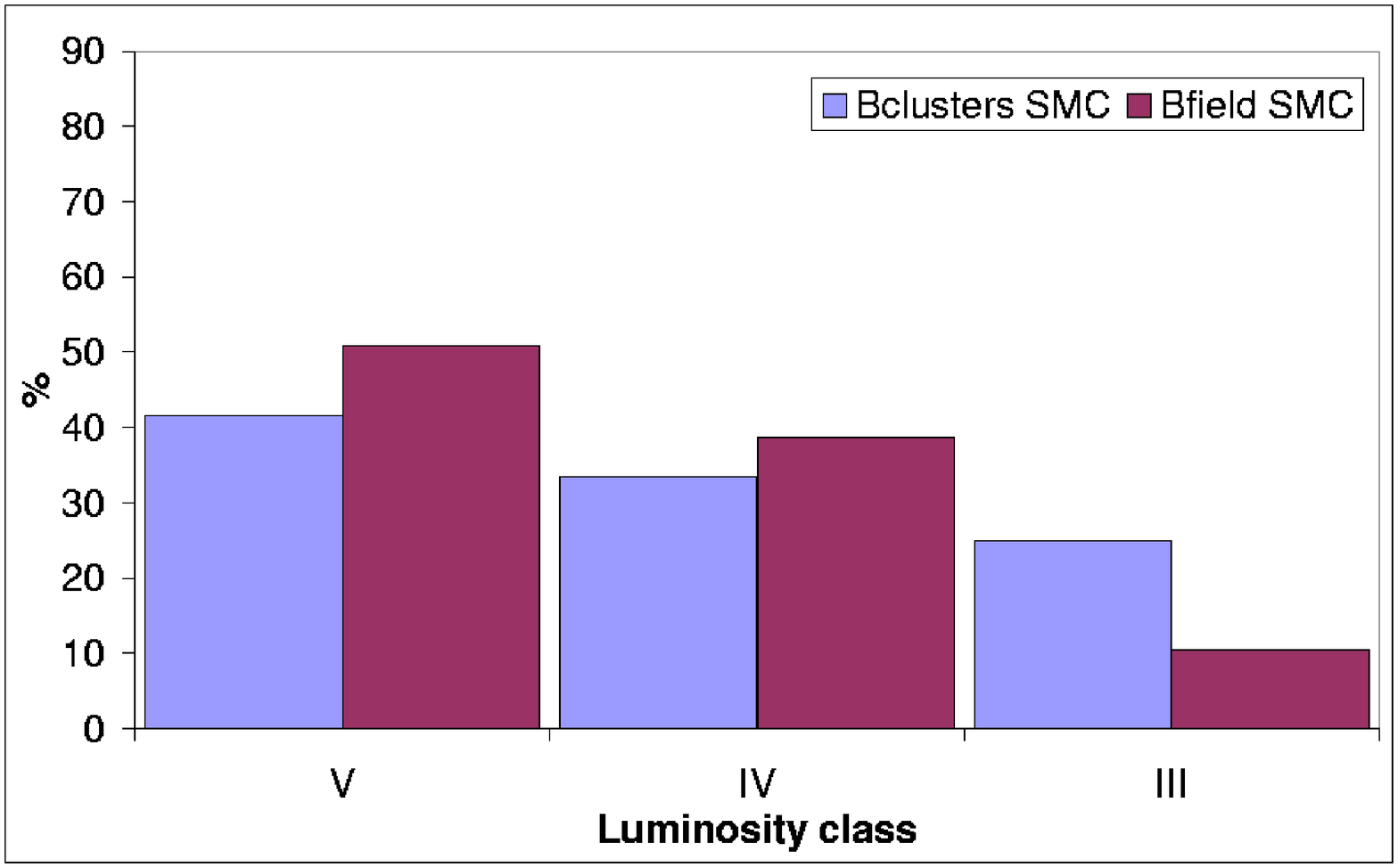}} 
\caption{Spectral type (upper panel) and luminosity class (lower panel)
distributions of B-type stars in the sample in the SMC.
In common: the blue left bars are for stars in clusters and the red right bars are for stars in fields.}
\label{STbappMC}
\end{figure}

We present in Fig.~\ref{STbappMC} the distribution of O-B-A stars with respect to
spectral type and luminosity class. The classification used here is the one obtained
from the fundamental parameters determination (CFP determination).

The sample contains essentially early B-type stars (B0 to B3) as in
the LMC (Paper I) which are mainly dwarfs and subgiants (classes V,
IV), in the field as well as in clusters.

\begin{figure}[!htbp]
\centering
\resizebox{\hsize}{!}{\includegraphics[angle=-90]{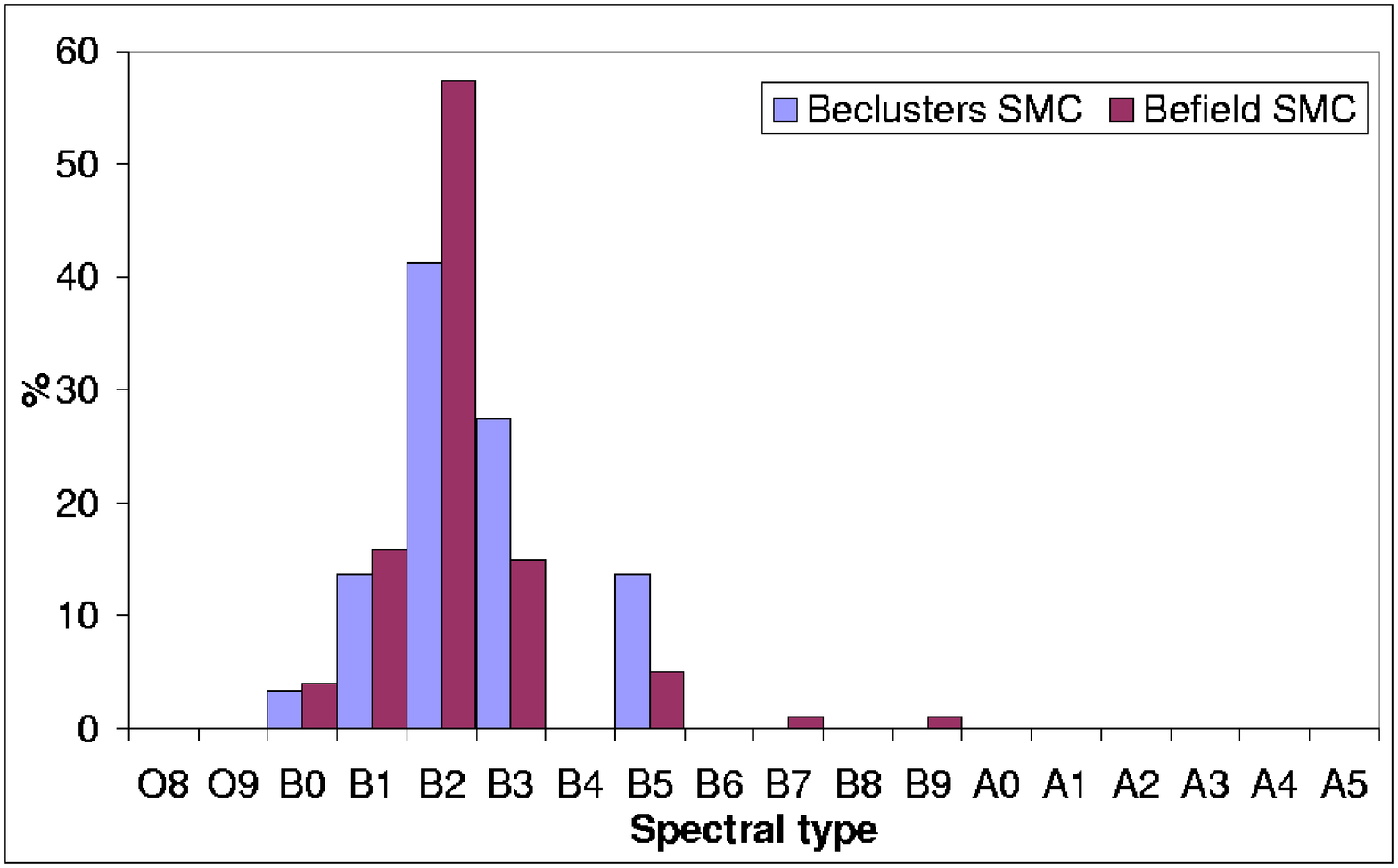}}\\
\resizebox{\hsize}{!}{\includegraphics[angle=-90]{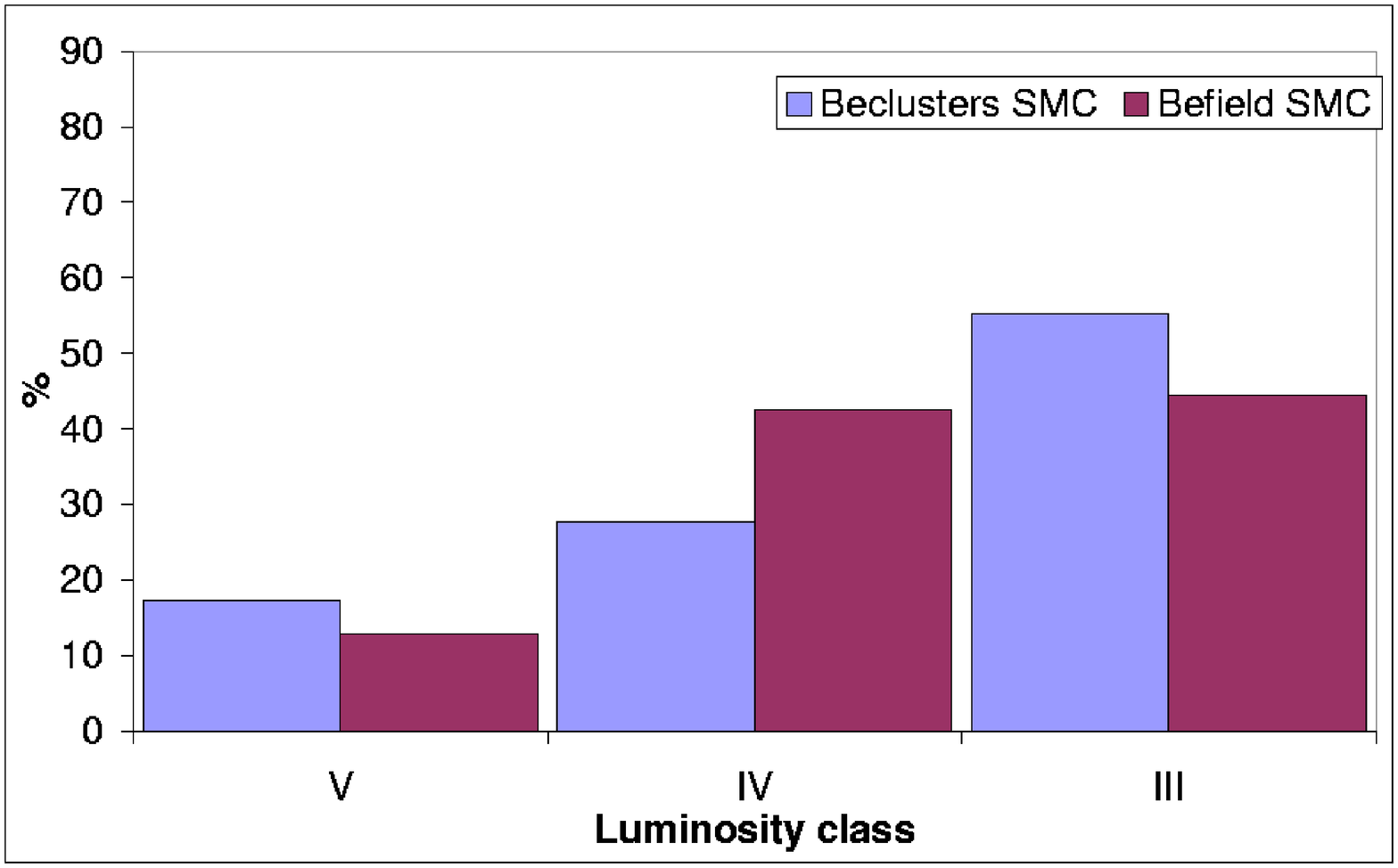}} 
\caption{Apparent spectral type (upper panel) and luminosity class (lower panel)
distributions of Be stars in the sample in the SMC.
In common: The blue left bars are for stars in clusters and the red right bars are for stars in fields.}
\label{STLCbeMC}
\end{figure}

\begin{figure}[!htbp]
\centering
\resizebox{\hsize}{!}{\includegraphics[angle=-90]{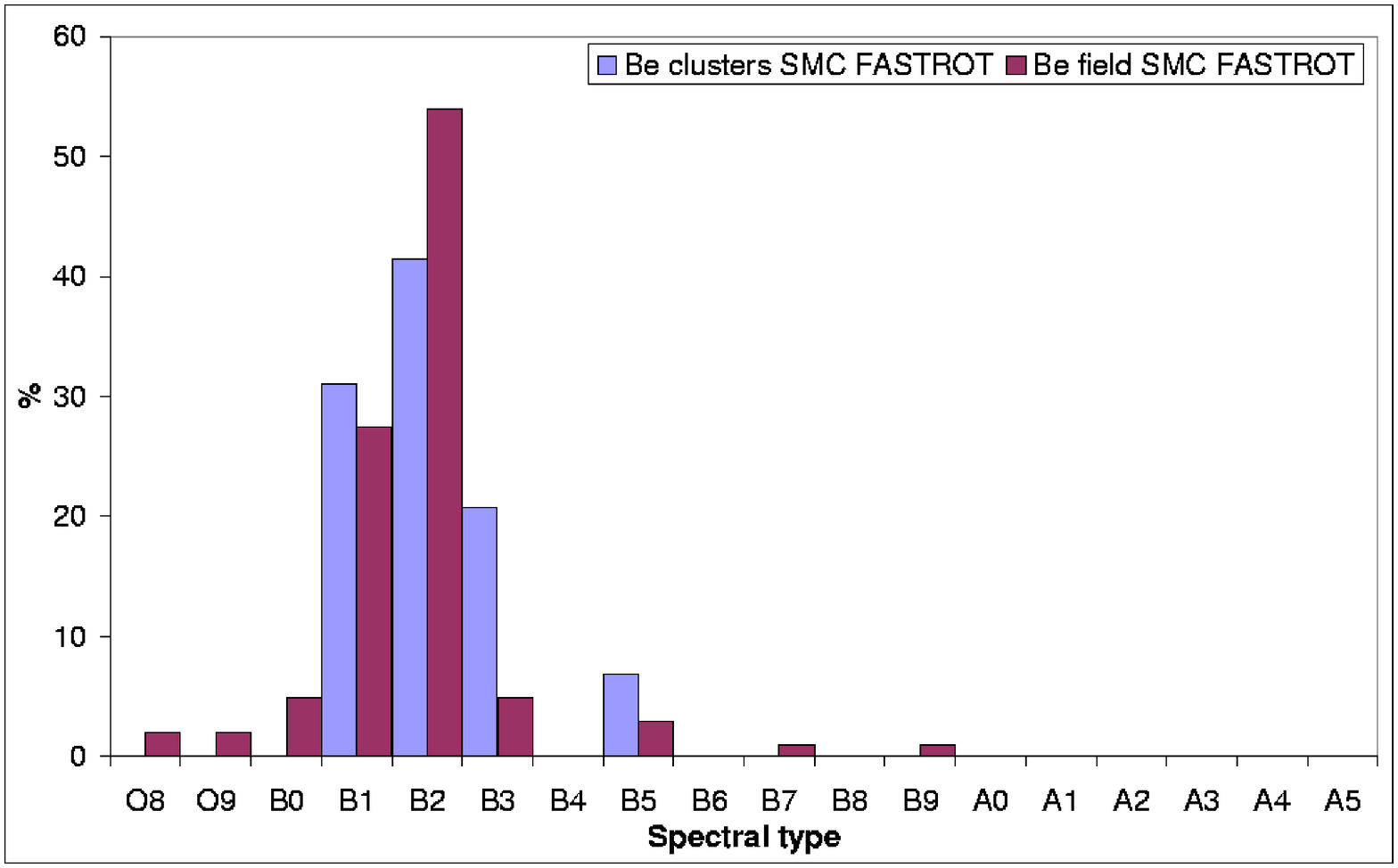}}\\
\resizebox{\hsize}{!}{\includegraphics[angle=-90]{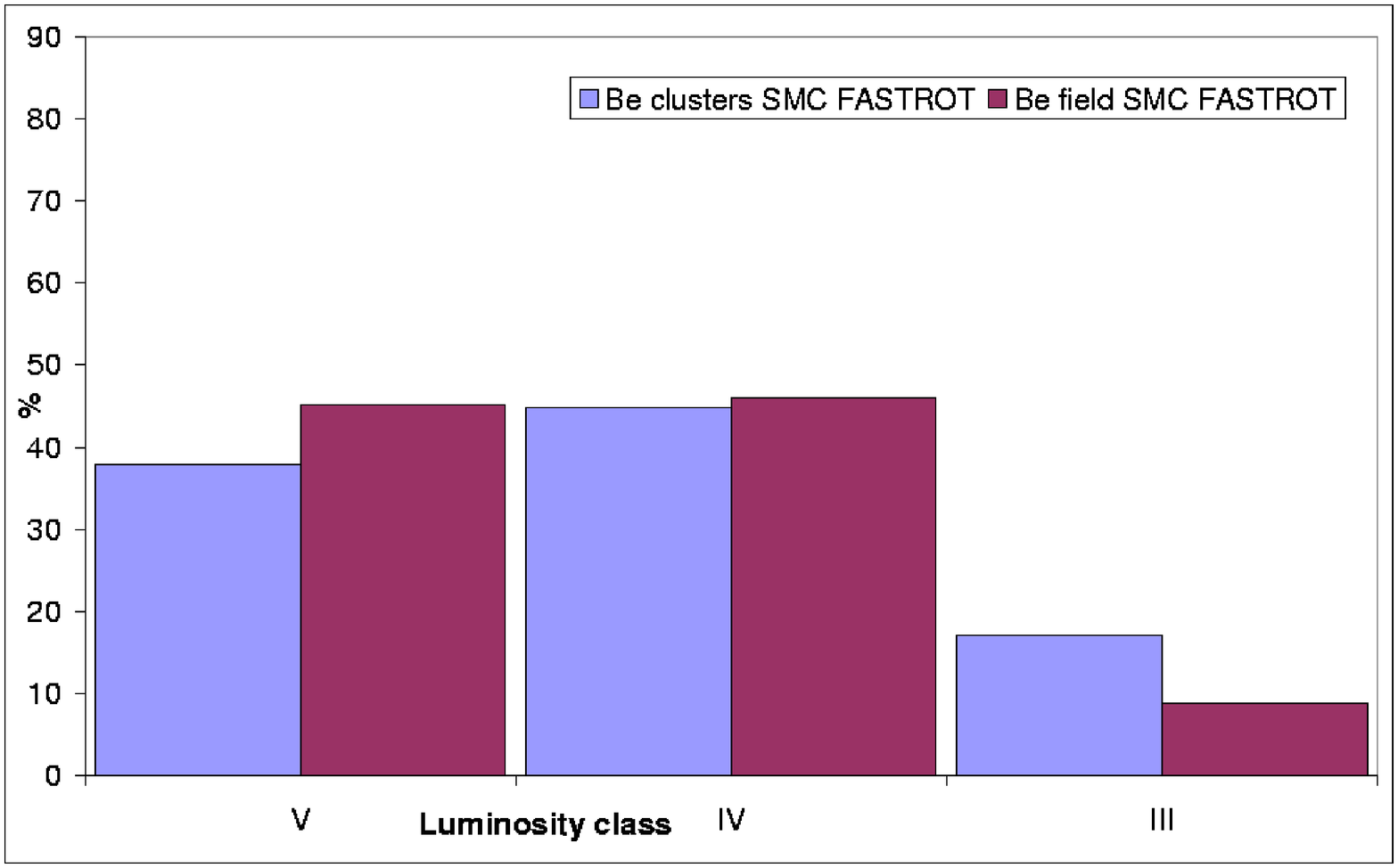}}
\caption{Corrected spectral type (upper panel) and luminosity class (lower panel)
distributions of Be stars after fast rotation treatment in the sample in the SMC.
In common: The blue left bars are for stars in clusters and the red right bars are for stars in fields.}
\label{STLCbeMCfast}
\end{figure}

We also present the distribution of Be stars with
respect to luminosity class and spectral type, using the 
classification obtained from the fundamental parameters. We compare the
distribution obtained before and after correction of fast rotation effects
(Figs.~\ref{STLCbeMC} and \ref{STLCbeMCfast}).


As for B stars, Be stars in our sample generally are early B-type stars (B0 to
B3) but are apparently giants and subgiants (classes III, IV). The
Be stars corrected for rotation effects appear hotter than apparent
fundamental parameters would suggest. In particular, there are more 
B1-type stars. After fast rotation treatment Be stars in classes III and IV are
redistributed in classes IV and V. However, about 60\% of the Be stars still appear
as giants and subgiants as in the LMC (Paper I).

\subsubsection{Distribution in masses}
\label{masses}

\begin{figure}[!ht]
%\centering
\resizebox{\hsize}{!}{\includegraphics[angle=-90]{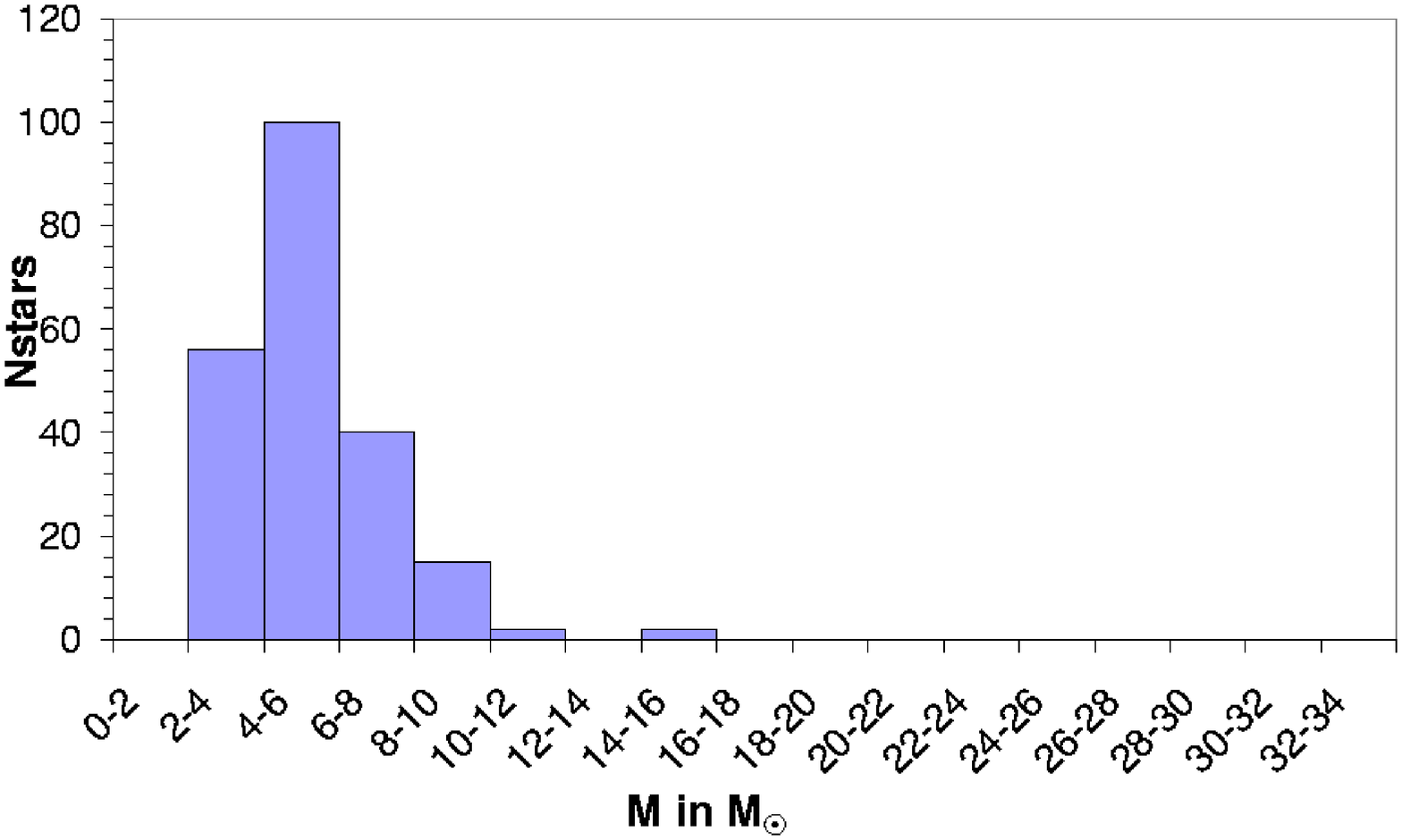}}\\
\resizebox{\hsize}{!}{\includegraphics[angle=-90]{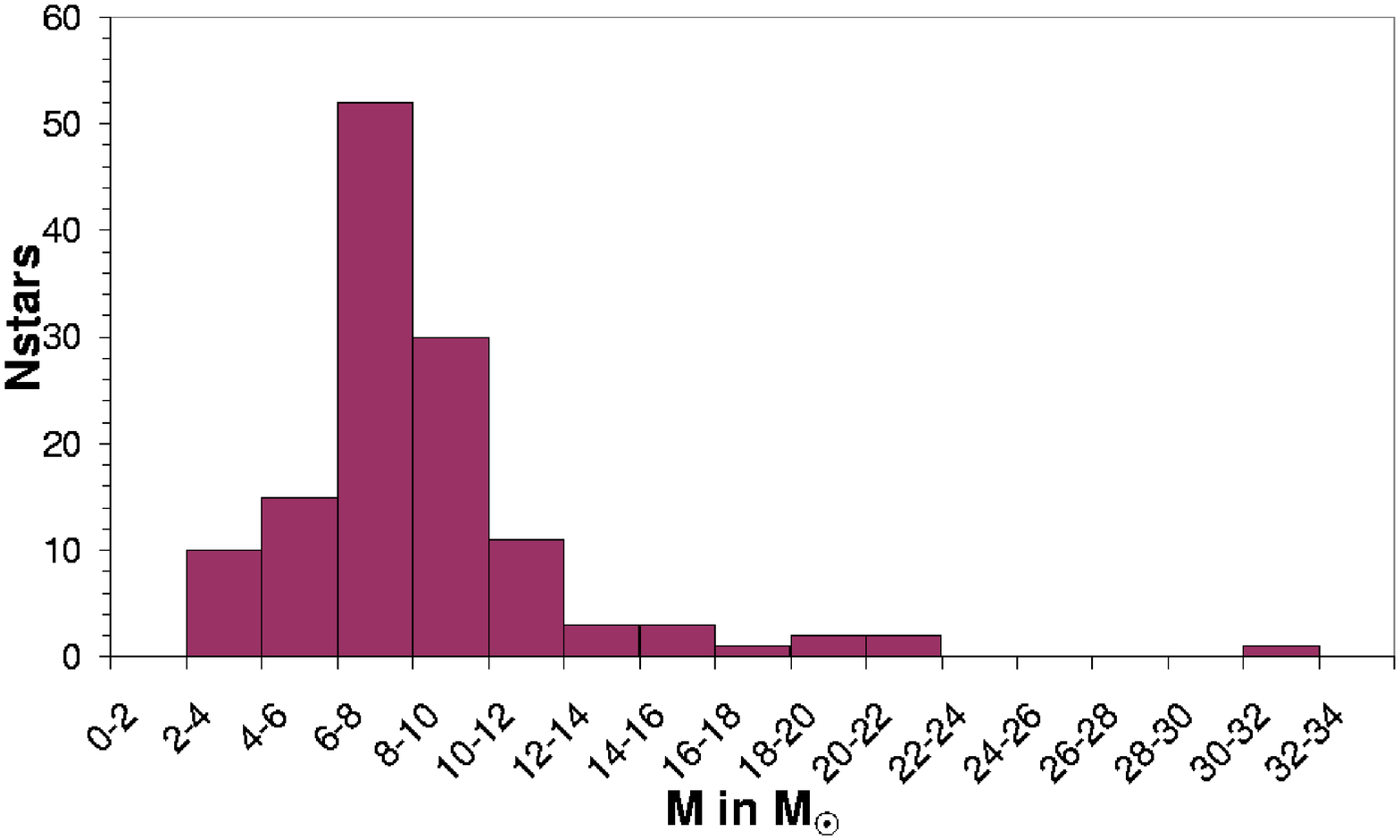}} 
\caption{Mass distribution of B (upper panel) and apparent mass distribution 
of Be stars (lower panel) in the sample in the SMC.}
\label{massesBBe}
\end{figure}

In addition, we investigate the mass distribution of B and Be stars
(Fig.~\ref{massesBBe}). The sample shows a distribution peaking around 5-6 and
7-8 M$_{\odot}$ for B and Be stars, respectively. These peaks are reminiscent of
those of B and Be stars in LMC's sample (7 and 10 M$_{\odot}$) as shown in Paper
I, but are shifted to smaller masses.

\subsubsection{Ages of clusters}
\label{agescl}

We determine the ages of stars of the field and of several clusters or
associations in our observations. For this purpose, we use HR evolutionary
tracks (for non-rotating stars) for the stars of the sample  unaffected by rapid
rotation and for Be stars corrected for the effects of fast rotation with
\omc~=~95\%. For the cluster NGC\,330, we obtain $\log(t)$ = 7.5$\pm$0.2, which
is  in excellent agreement with the value found photometrically by OGLE:
7.5$\pm$0.1 (see Pietrzy\'nski \& Udalski 1999) while Chiosi et al. (2006) 
found $\log(t)$ = 8.0. 
For the clusters OGLE-SMC99 and
OGLE-SMC109, we find $\log(t)$ = 7.8$\pm$0.2 and $\log(t)$ = 7.9$\pm$0.2 to be
compared  with :$\log(t)$ = 7.6$\pm$0.2 and $\log(t)$ = 7.7$\pm$0.1, respectively
(Pietrzy\'nski \& Udalski 1999), with 7.3 and 7.4 respectively from Chiosi et al. (2006), and with 
8.1 and 7.8 respectively from Rafelski \& Zaritsky (2005) for a metallicity Z=0.001. 
In the same way, we determine the age of other
clusters: for NGC299 $\log(t)$ = 7.8$\pm$0.2 and NGC306 $\log(t)$ = 7.9$\pm$0.2  to 
be compared with the values from Rafelski \& Zaritsky (2005): 7.9 and 8.5 respectively.
Our values are thus in general in good agreement with OGLE
and other determinations.
As for the LMC, these comparisons validate our method to
determine ages for clusters.

\section{Rotational velocity and metallicity: results and discussion}

We compare the \vsini~values obtained for B and Be stars in the SMC to
those obtained for the LMC (Paper I) and given in the literature for
the MW. However, the later generally did not take fast rotation effects into
account in the determination of fundamental parameters. Therefore, to allow the
comparison with the MW, we report on the apparent rotational velocity in
the case of rapid rotators in the SMC and LMC. 

\subsection{Vsini for the SMC in comparison with the LMC and MW}

For the same reasons as in Paper I, we cannot directly compare the
mean \vsini~values of the sample in the SMC with values in the LMC and in the
MW, because they are affected by ages and evolution, mass function of samples,
etc. We must therefore select B and Be stars in the same range of spectral types
and luminosity classes or of masses (when they are known) and ages for samples
in the SMC, LMC, and MW.

To investigate the effect of metallicity and age on the rotational velocity,
we first compare the mean \vsini~of B and Be stars either in the field or in clusters
in the SMC to the ones in the LMC and MW. Then, we compare the rotational velocity of
B and Be stars in field versus clusters in the SMC. We use the same selection criteria
as those for the LMC and MW described in  Paper I: we select stars with spectral type
ranging from B1 to B3 and luminosity classes from V to III.  For reference studies in
the LMC and MW, we use the same as those mentioned in Paper I (see Table 10
therein). The values are reported in Table~\ref{Vsinifieldcl}. As in Paper I, 
we recall that all suspected binaries, shown in Martayan et al. (2005b)
are removed from the statistics of the following sections. 

\begin{table*}
\caption{Comparison of mean rotational velocities for B and Be stars with
spectral types B1-B3 and luminosity classes from V to III in the SMC, LMC and MW. 
The values in brackets represent the number of stars in the samples.}
\centering
\begin{tabular}{lllllll}
\hline
\hline
From &                  &  Field B stars          &  Field Be stars         & Clusters B stars     & Clusters Be stars \\
\hline
this study & SMC  & 159 $\pm$ 20 (147) & 318 $\pm$ 30 (87) & 163 $\pm$ 18 (19) & 264 $\pm$ 30 (25) \\
\hline
Paper I & LMC  & 121 $\pm$ 10 (81) & 268 $\pm$ 30 (26)  &144 $\pm$ 20 (10) & 266$ \pm$ 30  (19)\\
Paper I & LMC Keller (2004) & 112 $\pm$ 50 (51) &                              & 146 $\pm$ 50 (49) &                     \\
\hline
Paper I & MW Glebocki et al. (2000) & 124 $\pm$ 10 (449)&   204 $\pm$ 20 (48)   &                             &                          \\
Paper I & MW Levato et al. (2004) & 108 $\pm$ 10 (150) &                                &                            &                          \\
Paper I & MW Yudin (2001) &                                 &  207 $\pm$ 30 (254)  &                           &                          \\
Paper I & MW Chauville et al. (2001) &                 &  231 $\pm$ 20 (56)  &                     &                                 \\
\hline
Paper I & MW WEBDA $\log(t)$ $<$ 7 &                    &                                     & 127 $\pm$ 20 (44)  & 199 $\pm$ 20 (8)     \\
Paper I & MW WEBDA $\log(t)$ $\ge$ 7 &                     &                                     & 149 $\pm$ 20 (59)  &   208 $\pm$ 20 (45)   \\
\hline
\end{tabular}
\label{Vsinifieldcl}
\end{table*}
\begin{figure}[]
\centering
\resizebox{\hsize}{!}{\includegraphics[angle=-90]{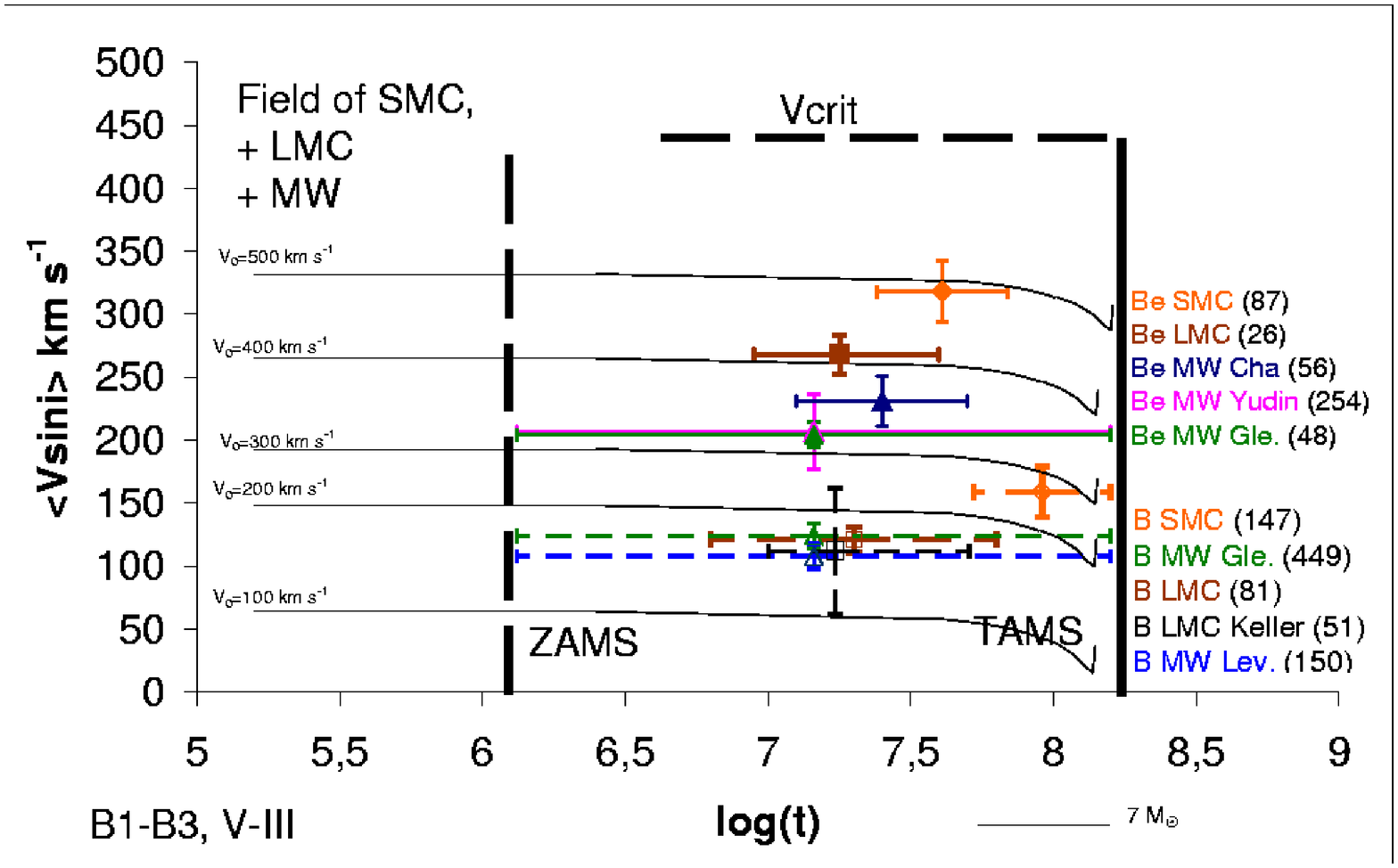}}\\
\resizebox{\hsize}{!}{\includegraphics[angle=-90]{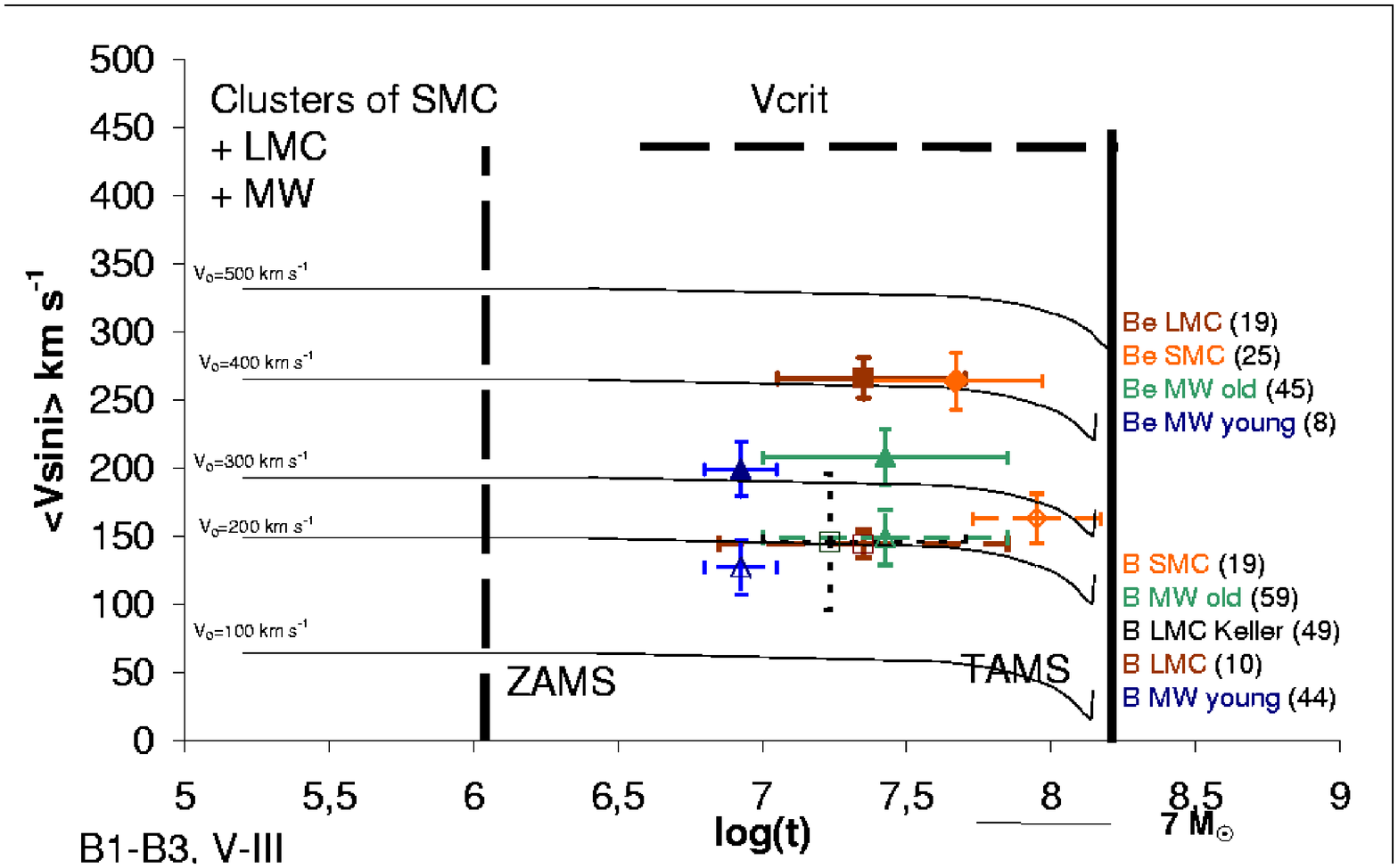}}
\caption{Comparison of mean \vsini~in the SMC, in the LMC and in the MW.
 Evolutionary tracks of rotational velocity during the main sequence life
are given for different initial velocities for a 7 M$_{\odot}$ star. The ZAMS
and TAMS are indicated by vertical lines and the critical \vsini~by a
horizontal dotted line. The number of stars for each study is given in
brackets. The dispersion in ages corresponds to the range of individual stellar ages
in the samples, when these ages are known, or to the main sequence lifetime.
Upper panel: for field B and Be stars.
The considered studies are: for the SMC, this paper; for the LMC, Martayan et al. (2006b) 
and Keller (2004); for the MW, Cha = Chauville et al. (2001), Yudin (2001), 
Gle = Glebocki \& Stawikowski (2000), and Lev = Levato \& Grosso (2004).
Lower panel: Same figure but for clusters. The considered
studies are: for the SMC, this paper; for the LMC, Paper I and Keller (2004), 
and for the clusters in the MW, the WEBDA database.}
\label{vsiniselectfieldclusters}
\end{figure}

\subsubsection{Field B and Be stars}
\label{BBefield}

The comparison of \vsini~in the SMC, the LMC and the MW for B and Be stars in the
field are presented in Table~\ref{Vsinifieldcl} and
Fig.~\ref{vsiniselectfieldclusters} (upper panel). In the figure the range of stellar
ages is reported as the dispersion in age. For samples with an unknown age, we adopt
as error bar the duration of the main sequence for a 7 M$_{\odot}$ star, which
highly overestimates the age uncertainty. The curves show the evolutionary tracks
of rotational velocity during the main sequence for different initial velocities for a 7
M$_{\odot}$ star, which corresponds to the maximum of the mass function of the B-star
sample. These curves are obtained as described in Paper  I (Sect. 5.2). 

As for the LMC, we show that the samples in the SMC contain a sufficient
number of elements for the statistics to be relevant and give an average \vsini~not
biased by inclination effects. We complete the statistical study by the Student's
t-test (Table~online) in order to know whether the differences observed
between samples in the SMC, LMC and MW are significant. \\
(i) for field B stars: we find that there is a significant difference between the SMC,
the LMC, and the MW. Field B stars in the SMC have a rotational velocity higher than
in the LMC and the MW. We recall that the test is not conclusive between B stars in
the LMC and the MW (Paper I), because results are different according to
the selected study in the MW (Glebocki et al. 2000 or Levato et al. 2004).\\
(ii) for field Be stars: there is a slight difference between the SMC and
the LMC, and a significant difference between the SMC and the MW. Field Be stars in
the SMC have a higher rotational velocity than in the LMC and the MW. We
also recall that, from Paper I, field  Be stars in the LMC have a
rotational velocity higher than in the MW.

\subsubsection{B and Be stars in clusters}
\label{BBecl}

The comparison of \vsini~in the SMC, LMC and MW for B and Be stars in clusters are
presented in Table~\ref{Vsinifieldcl} and Fig.~\ref{vsiniselectfieldclusters} (lower
panel). The evolutionary tracks curves are the same as in the upper panel.  For the MW, we use the
selection we made in Paper I. We  distinguish two groups: the younger
clusters with $\log(t)$ $<$ 7 and older clusters with $\log(t)$ $\ge$ 7. The
age-difference between clusters, taken from WEBDA, gives the age-dispersion reported
in the figure. The results concerning B and Be stars in the SMC, LMC and MW
clusters are:\\
(i) B stars in the SMC and LMC clusters seem to have a similar rotational
velocity, as in the MW when interval of similar ages are compared (Paper I).\\
(ii) for Be stars: We note no difference between Be stars in the SMC and LMC clusters,
while there is a significant difference between the LMC and the MW clusters. Be stars
rotate more rapidly in the Magellanic Clouds (MC) clusters than the MW clusters. The
lack of difference between the SMC and the LMC is probably due to a
difference in mass and evolution functions of the stars in the samples (see 
 Section~\ref{masses} and Paper I, Sect. 4.5.3).

\subsubsection{Comparison between field and clusters}

No significant differences can be found between rotational velocities  neither for 
field versus cluster B stars in the SMC, LMC, and MW, nor for field versus cluster Be
stars in the LMC and the MW. However, a slight trend seems to be present for Be stars
in the SMC. Field Be stars seem to rotate faster than cluster Be stars in the SMC. 
However, note the large error bar on the mean \vsini~value for Be stars in the SMC,
which prevents conclusive results between field and clusters.

\subsection{B and Be stars: mass and rotation}
\label{BBemasses}

The search for links between metallicity and rotation of B and Be stars is also carried
out thanks to a selection by masses, which allow a direct comparison with theoretical
tracks of the rotational velocities. To obtain sub-samples in the most homogeneous
possible way, we select the stars by mass categories: 5 $\leq$ M $<$ 10 M$_{\odot}$, 
10 $\leq$ M $<$ 12M$_{\odot}$, etc.  We assume a random distribution for the
inclination angle. 

The number of observed stars for a given mass category is low, therefore we have not
separated the stars in clusters and in fields categories. This is justified
since we have not found any significant difference between the rotational velocities
of stars in fields and in clusters. We first present a general result between B and Be
stars in the MC, then we study in detail the effects of metallicity and evolution on
the rotational velocities for Be stars in the MW and in the MC.

\begin{figure*}[!htbp]
\centering
\resizebox{\hsize}{!}{\includegraphics[angle=-90]{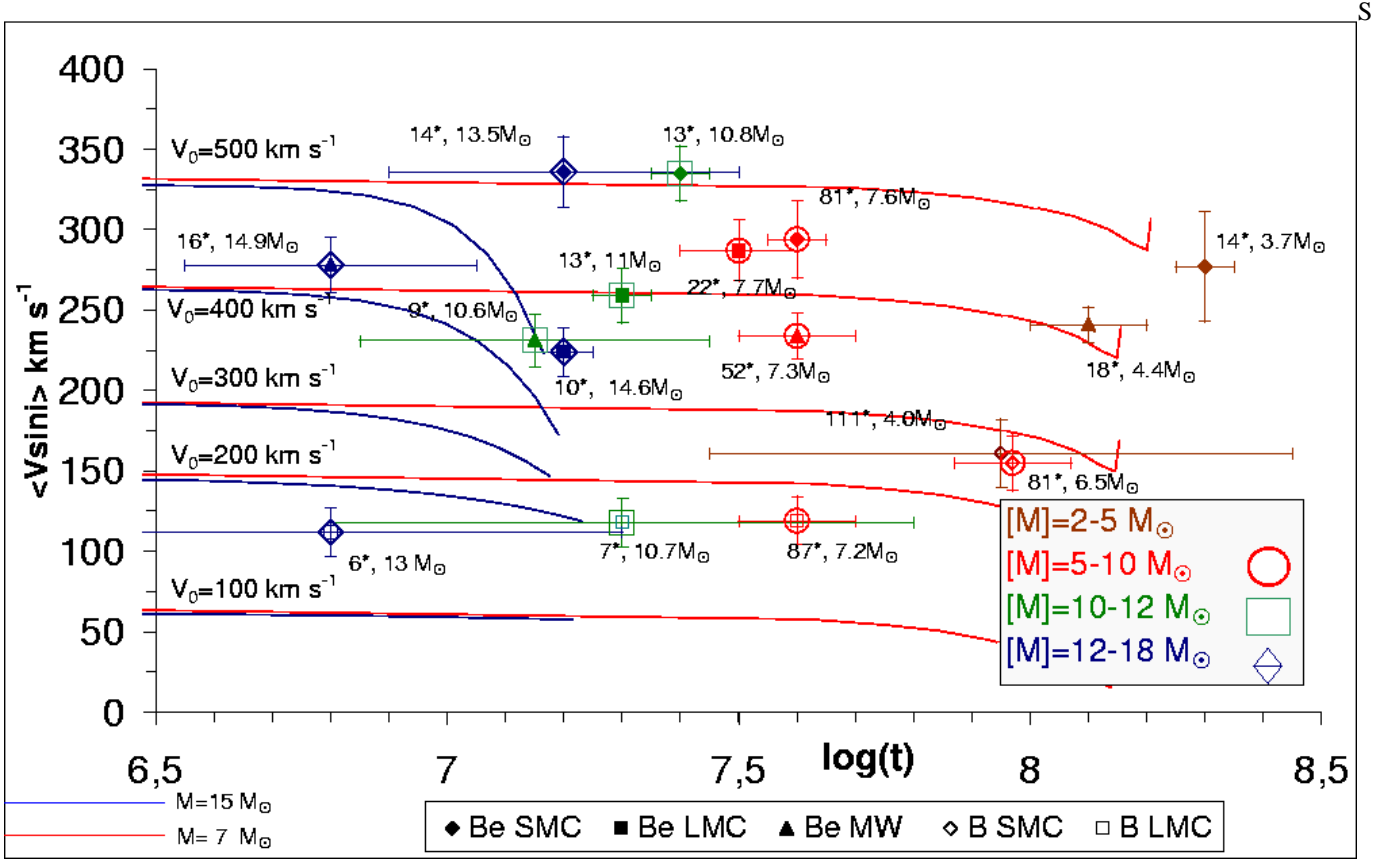}}S
\caption{Comparison of the rotational velocities for Be stars in the SMC, in the LMC and in the MW; 
and for B stars in the SMC and the LMC.
The squares are for the samples of stars in the LMC, the diamonds are for the samples of stars in the SMC, 
and the triangles are for the samples of stars in the MW. Empty symbols are for the B stars, and full symbols
are for Be stars.
The different mass-categories are indicated by different colours and symbols which surround the symbols 
for the B and Be stars in the 3 galaxies:
brown for masses ranging from 2 to 5 M$_{\odot}$, 
red large circles for masses ranging from 5 to 10 M$_{\odot}$, 
green large squares for masses ranging from 10 to 12 M$_{\odot}$, 
and blue large diamonds for masses ranging from 12 to 18 M$_{\odot}$.
The numbers indicated next to each point correspond to the number of stars '*' in each sample and to their mean mass.
The tracks of rotational velocities for a 7 and a 15 M$_{\odot}$ 
star obtained for the SMC from our interpolations in the studies of Meynet \& Maeder (2000, 2002), 
Maeder \& Meynet (2001) are shown to illustrate the results.} 
\label{compBBeMCtout}
\end{figure*}

The rotational velocities we derived from observational results, for the different samples
by mass categories for B and Be stars in the MC are reported in
Fig.~\ref{compBBeMCtout}. This graph shows, as from spectral type-selection, that Be stars
reach the main sequence with high rotational velocities at the ZAMS 
contrary to B stars. Consequently, only a B star with a sufficiently high initial rotational
velocity at the ZAMS may become a Be star.

\subsubsection{Effect of metallicity on B stars}

The values of the mean rotational velocities for the mass-samples of B stars in the SMC
and LMC are given in Table~\ref{vsiniBmasses} and reported in Fig.~\ref{compBBeMCtout}. For
a better comparison, we must reduce the number of degrees of freedom, i.e. compare the
samples with similar ages and similar masses and with a sufficient number of elements.
That is the case, in this study,  for the 5-10 M$_{\odot}$ category sample. 
Even if the age is not exactly the same, the evolution
of the rotational velocity as shown by the tracks for a 7 M$_{\odot}$, 
indicate that the velocity of the SMC's sample is higher than the one of the LMC at
exactly the same age. A 
highly significant difference (probability 99.5\%) in the mean rotational velocity is then 
shown for this mass-category between the SMC and the LMC with the Student's t-test. 
Note that such a comparison cannot
be done with B stars in the MW, because the data in the literature do not allow to
determine their masses.

\addtocounter{table}{+1}
\begin{table*}[th]

\caption[]{Comparison by mass sub-samples of the mean rotational velocities in the SMC and LMC B stars.
For each sub-sample, the mean age, mean mass, mean \vsini~and the number of stars (N*) are given.
No result is given for massive stars in the SMC, because of their small number.}
\centering
\begin{tabular}{l|cccc|cccc}
\hline
\multicolumn{1}{c}{}&\multicolumn{4}{c}{2-5 M$_{\odot}$}&\multicolumn{4}{c}{5-10 M$_{\odot}$}\\
\hline
	& $<$age$>$ & $<$M/M$_{\odot}$$>$& $<$\vsini$>$ & N* & $<$age$>$ & $<$M/M$_{\odot}$$>$& $<$\vsini$>$ & N* \\
\hline
SMC B stars & 8.0 & 4.0 & 161 $\pm$ 20 & 111 & 8.0 &  6.5 & 155 $\pm$ 17 & 81 \\
LMC B stars & 7.9 & 4.1 & 144 $\pm$ 13 &   6  & 7.6 & 7.2 & 119 $\pm$ 11 & 87 \\
\hline

\end{tabular}

\begin{tabular}{l|cccc|cccc}
\multicolumn{1}{c}{}&\multicolumn{4}{c}{10-12 M$_{\odot}$}&\multicolumn{4}{c}{12-18 M$_{\odot}$}\\
\hline
	& $<$age$>$ & $<$M/M$_{\odot}$$>$ & $<$\vsini$>$ & N*& $<$age$>$ & $<$M/M$_{\odot}$$>$ & $<$\vsini$>$ & N* \\
\hline
SMC B stars &  &  &  &  3 &  &  &  &  2 \\
LMC B stars & 7.3 & 10.7 & 118 $\pm$ 10 &  7  & 6.8 & 13.0 & 112 $\pm$ 10 &  6 \\
\hline

\end{tabular}
\label{vsiniBmasses}

\end{table*}

The effect of metallicity on rotational velocities for B stars is
shown for the 5-10 M$_{\odot}$ category with similar ages between the SMC and 
the LMC: the lower the metallicity, the higher the rotational velocities of B
stars. It could also be valid for other ranges of masses, but this has to be
confirmed with appropriate samples. Nevertheless, this  observational result
nicely confirms the theoretical result of Meynet \& Maeder (2000) and Maeder \&
Meynet (2001).

\subsubsection{Effect of metallicity on Be stars}

The selection by mass samples of Be stars  in the SMC was done as in
Paper I for the LMC. For the MW, we use the studies published by Chauville et
al. (2001) and Zorec et al. (2005). The mean rotational velocity for each mass
sample of Be stars is reported in Table~\ref{vsiniBemasses} and is shown in
Fig.~\ref{compBBeMCtout} for the three galaxies. For each sample, the mean age
and mass, as well as the number of considered stars are given.

\begin{table*}[!th]
\caption{Comparison by mass sub-samples of the mean rotational velocities for the samples of Be stars in the SMC, LMC and in the MW.
For each sample, the mean age, the mean mass, the mean rotational velocity and the number of stars are given.
Note no low-mass Be star in the LMC.}
\centering
\begin{tabular}{l|cccc|cccc}
\hline
\multicolumn{1}{c}{}&\multicolumn{4}{c}{2-5 M$_{\odot}$}&\multicolumn{4}{c}{5-10 M$_{\odot}$}\\
\hline
	& $<$age$>$ &  $<$M/M$_{\odot}$$>$ & $<$\vsini$>$ & N* & $<$age$>$ & $<$M/M$_{\odot}$$>$ & $<$\vsini$>$ & N* \\
\hline
SMC Be stars & 8.0 &  3.7 & 277 $\pm$ 34 & 14 & 7.6 & 7.6 & 297 $\pm$ 25 & 81 \\
LMC Be stars &  & & &  0 & 7.5 & 7.7 & 285 $\pm$ 20 & 21 \\
MW Be stars & 8.1 & 4.4 & 241 $\pm$ 11 & 18 & 7.6 & 7.3 & 234 $\pm$ 14 & 52 \\ 
\hline
\end{tabular}

\begin{tabular}{l|cccc|cccc}
\multicolumn{1}{c}{}&\multicolumn{4}{c}{10-12 M$_{\odot}$}&\multicolumn{4}{c}{12-18 M$_{\odot}$}\\
\hline
	& $<$age$>$ & $<$M/M$_{\odot}$$>$ & $<$\vsini$>$ & N*& $<$age$>$ & $<$M/M$_{\odot}$$>$ & $<$\vsini$>$ & N* \\
\hline
SMC Be stars & 7.4 &  10.8 &  335 $\pm$ 20 &  13 & 7.2 & 13.5 & 336 $\pm$ 40 &  14\\
LMC Be stars & 7.3 & 11 & 259 $\pm$ 20 & 13 & 7.2 & 14.6 & 224 $\pm$ 30 & 10\\
MW Be stars & 7.2 & 10.6 & 231 $\pm$ 16 & 9 &  6.8 & 14.9 & 278 $\pm$ 10 &  17\\
\hline
\end{tabular}

\label{vsiniBemasses}
\end{table*}

From the results of the Student's t-test in Table~online, 
we conclude that there is an effect of metallicity on the
rotational velocities for Be stars in samples with similar masses and similar ages: the
lower the metallicity, the higher the rotational velocities. 
This is particularly visible in high-mass (10-12 M$_{\odot}$) and intermediate-mass
(5-10 M$_{\odot}$) samples of Be stars in the SMC and the MW.

Note, moreover, that for the most massive-star samples (12 $\leq$ M $<$ 18 M$_{\odot}$), it
is more difficult to compare them directly, since the ages are quite different between the
MC and MW.  We note the lack of massive Be stars in the MW at ages for which Be stars are
found in the MC. It suggests that the Be star phase can last longer in low metallicity
environments such as the MC, compared to the MW.

\subsection{ZAMS rotational velocities of Be stars}
\label{ZAMSRvot}

The interpretation of our results requires a set of rotational velocity tracks for masses
between 2 and 20 M$_{\odot}$, for different metallicities corresponding to the MC and the
MW, and for different initial rotational velocities at the ZAMS. We obtain these tracks by
interpolation in the models of the Geneva group as described in 
Sect.~\ref{theotracks}. For the first time we derive the distributions of the ZAMS
rotational velocities of Be stars as shown in Sect.~\ref{ZAMSrotBe}.

\subsubsection{Theoretical evolutionary tracks of the rotational velocity}
\label{theotracks}

We derive rotational velocity  evolutionary tracks for different masses, initial
velocities and metallicities by interpolation in curves published by Meynet \&
Maeder (2000, 2002), Maeder \& Meynet (2001). We proceed 
as reported in Paper I (Section 5.2) for a 7 M$_{\odot}$ star.  We then  obtain  
curves  for 3, 5, 7, 9, 12, 15, and 20 M$_{\odot}$ stars, for initial rotational velocities
V$_{0}$=100, 200, 300, 400, 500 \kms, and at available metallicities Z=0.020 (solar
metallicity), Z=0.004 (LMC), and Z=0.00001 (metallicity similar to the one of the first
generation of stars). For Z=0.001 (SMC), the curves result from our interpolations.
The increase in the lifetime of stars on the main sequence due to
rotation and metallicity is taken into account.

Note that due to fast internal angular momentum redistribution in the first
$\simeq$10$^{4}$ years in the ZAMS, the surface rotational velocities decrease 
by 0.8 to 0.9 times their initial value. Moreover, for the comparison-sake with our
observational data, the values plotted are not $V$ but are averaged \vsini$=(\pi/4)V$. For
example, for an initial rotational velocity equals to 300 \kms, the angular momentum
redistribution leads roughly to V$_{ZAMS}=$ 240 \kms, which corresponds to \vsini$=(\pi/4)
\times 240$ $\simeq$ 190 \kms. 

\begin{figure}[!htbp]
\centering
\resizebox{\hsize}{!}{\includegraphics[angle=-90]{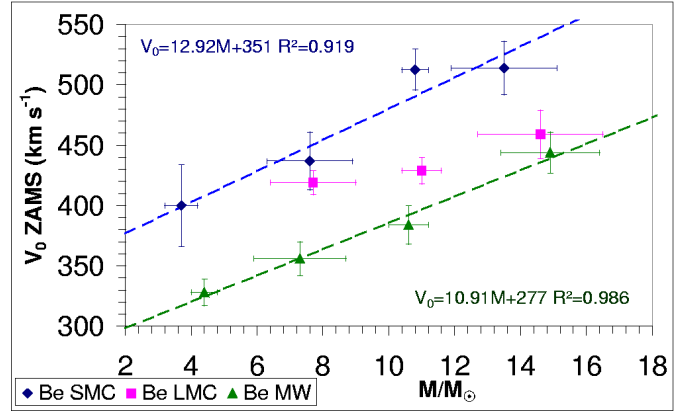}}
\caption{ZAMS rotational velocities for the samples of Be stars in the SMC (blue diamonds), in the LMC (pink squares)
and in the MW (green triangles). The dashed lines correspond to the linear regressions. 
Their corresponding equations and correlation coefficient are given in the upper left and lower right corners.
}
\label{V0ZAMSequawind}
\end{figure}

\subsubsection{ZAMS rotational velocities}
\label{ZAMSrotBe}

Our study shows that Be stars begin their life on the main sequence with higher
rotational velocities than those for B stars. For each sample of Be stars in the
SMC, LMC and  MW, the initial rotational velocity at the ZAMS has been obtained
by interpolation between the tracks of the evolution of rotational velocity
during the MS (see Sect.~\ref{theotracks}). The resulting
distributions of ZAMS rotational velocities for the samples of Be stars in the
SMC, LMC and MW are shown in Fig.~\ref{V0ZAMSequawind}.

The average rotational velocities at the ZAMS (V$_{O}$) are quantified by linear regressions
and their equations are as follows:
\begin{itemize}	
\item In the SMC, V$_{O}$= 12.92 $\frac{M}{M_{\odot}}$ + 351 and the correlation
coefficient is R$^{2}$=0.919.
\item In the MW, V$_{O}$= 10.91 $\frac{M}{M_{\odot}}$ + 277 and the correlation
coefficient is R$^{2}$=0.986.
\item For the LMC, the lack of low mass stars in our sample implies that we cannot 
determine the ZAMS distribution. The later seems to be situated between the distributions of the SMC and MW.
\end{itemize}

\subsubsection{Consequences}

\begin{itemize}
\item Whatever the metallicity, the ZAMS rotational velocities of Be stars
depend on their masses.
\item Following the SMC and MW curves, the trend (the gradient) of the ZAMS rotational
velocities of Be stars could be independent of the metallicity.
\item There is an effect of metallicity on the distributions of the ZAMS rotational
velocities. The lower the metallicity, the higher the ZAMS rotational
velocities.
\item There is a limit of the ZAMS rotational velocities below which B stars
will never become Be stars. This limit depends on the metallicity.
\end{itemize}

The initial conditions (magnetic field, accretion disk, etc) in an open cluster will
lead to a more or less high number of B stars with a ZAMS rotational velocity high enough to become
Be stars. Therefore, the rates of Be stars will fluctuate depending on the cluster. It is
expected that the mean rate of Be stars at low metallicity, typically in the SMC, is higher
than at high metallicity, typically in the MW. This trend was already observed by Maeder
et al. (1999) in open clusters. We find a similar result between the fields in the LMC and
the SMC (Martayan et al. 2006c, in preparation). All these observational results give new
constraints on the pre-main sequence (PMS) evolution of B and Be stars progenitors and, in
particular, support the arguments exposed by Stepi\'en (2002) about the influence of
a magnetic field on formation conditions of B and Be stars (see Paper I, Sect. 
5.2.5). Progenitors of Be stars would possess a weak magnetic field with a surface intensity between
40 and 400 G and, due to the short PMS phase for the early types, would conserve their strong
rotational velocity during the main sequence. In low metallicity environments, the magnetic field has
less braking impact, which could explain why Be stars in the SMC can rotate initially with
higher velocities than in the LMC, and Be stars in the LMC with higher
velocities than in the MW, as shown in Figs.~\ref{vsiniselectfieldclusters}  and~\ref{compBBeMCtout}.
Note that a weak magnetic field is suspected in the classical Be star \object{$\omega$ Ori} (Neiner
et al. 2003).

\section{Angular velocity and metallicity: results and discussion}

Thanks to the formulae published in Chauville et al. (2001) and in Paper I, it
is possible to obtain the average ratio of angular to the breakup angular velocity for the
stars.

\subsection{Angular velocities for B stars}

We determine the mean \omc~ratio of B-type stars in the LMC (37\%) and in the SMC (58\%). 
In the same way, we determine this ratio in the MW with the values in Table~\ref{Vsinifieldcl}; 
this ratio ranges from 30 to 40\%.

The values of \omc seem to be similar for B stars in the MW and in
the LMC, but higher in the SMC. This difference is probably due to the large
difference of metallicity between the SMC and the LMC/MW. We recall that the considered stars 
in the MC have similar ages.

\subsection{Angular velocities for Be stars}

According to Porter (1996), the mean \omc~ratio of Be-type stars in the MW is 84\% in
good agreement  with the one (83\%) determined by Chauville et al. (2001). Following the
detailed study by Cranmer (2005) on Be stars in the MW, the ratio \omc~ranges from  69\% to
96\% for the early-types. In the LMC (Paper I) this ratio ranges from 73\% to 85\%,
and in the SMC (this study) from 94\% to 100\%. We recall that, in the MC, the observed Be stars
are also early-types and have similar ages. We thus note the
following trend: in the SMC they rotate faster than in the LMC/MW and are
close to the breakup velocity or are critical rotators. This shows that, in a low
metallicity environment such as the SMC, more massive Be stars can reach the critical
velocity. We also note that Be stars appear with at least \omc$\simeq$70\% in the LMC. This
value seems to be a threshold value to obtain a  Be star in the MW, LMC, and by extension
certainly in the SMC.

According to the stellar wind theories and Maeder \& Meynet (2001), the higher
the mass of the star, the higher the mass loss and angular 
momentum loss. However, in low metallicity environments, this mass loss and
consequently the angular momentum loss are lower than in the MW. As during the
main sequence the radius of the star increases and as the star conserves high
rotational velocities with a mass which decreases only slightly, the critical
velocity decreases. Consequently, the \omc~ratio for massive stars 
increases at low metallicity, while it decreases in the MW.  For intermediate
and low mass Be stars, whatever the metallicity, the \omc~ratio  
first stays relatively constant and then increases at the end of the 
main sequence.

\subsection{ZAMS angular velocities for Be stars}
\label{ZAMSomega}

Thanks to the distributions of ZAMS rotational velocities for Be stars presented
in Sect.~\ref{ZAMSRvot} and with the mass, and radius at the ZAMS from the Geneva models,
it is possible to obtain the ZAMS angular velocities and the \omc~ratio for Be stars 
in the MW and in the MC. 
The results are given in Table~\ref{omegaZAMSBe}. The effect of metallicity  on
the angular velocities is visible at the ZAMS. Note that the ratio
\omc~increases as the mass of the star increases.

All the theoretical calculations from Meynet \& Maeder (2000, 2002) and Maeder
\& Meynet (2001) have been performed with a ZAMS rotational velocity equal to
300 \kms. From the observations, we find that Be stars begin their MS lifetime with
ZAMS rotational velocities higher than 300 \kms. For example, for a 20 M$_{\odot}$ Be star,
the ZAMS rotational velocity is equal to 495 \kms~in the MW and 609
\kms~in the SMC. The ZAMS rotation rate for a B
star in the SMC is thus higher than the value adopted
by Meynet \& Maeder in their studies. In the SMC, it is consequently easier than expected 
for Be stars to reach the critical velocity. Note that, due to differential
rotation, these stars could be critical rotators at
their surface but not inside their core.

\addtocounter{table}{+1}
\begin{table}[th]
\caption{Ratio of angular velocity to the breakup angular velocity
\omc~(\%) at the ZAMS for Be stars in the MW and in the SMC (this study) for a 5, 12,
20 M$_{\odot}$ star.}
\centering
\begin{tabular}{@{\ }c@{\ \ \ }c@{\ \ \ }c@{\ \ \ }c@{\ \ \ }c@{\ \ \ }c@{\ \ \ }c@{\ \ \ }c@{\ \ \ }c@{\ \ \ }c@{\ }}
\hline
\hline
 & MW & SMC\\
\hline
5M$_{\odot}$ \omc (\%) & 75 & 77\\
12M$_{\odot}$ \omc (\%) & 80 & 84\\
20M$_{\odot}$ \omc (\%) & 85 & 92\\
\hline
\hline
\end{tabular}
\label{omegaZAMSBe}
\end{table}
\section{Evolutionary status of Be stars}
\subsection{Observational results for Be stars in the SMC}

\begin{figure}[!h]
\centering
\resizebox{\hsize}{!}{\includegraphics[angle=-90]{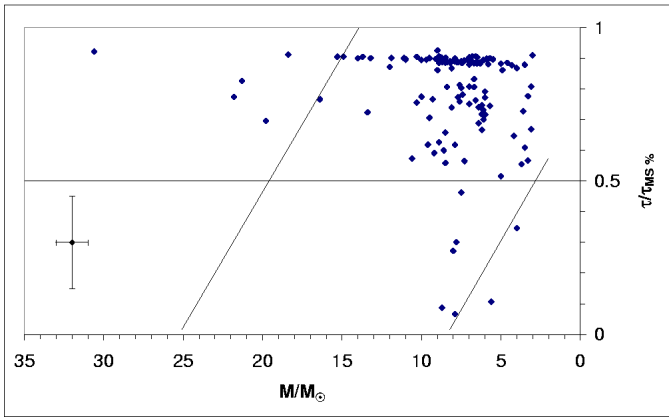}}
\caption{Evolutionary status of Be stars in the SMC. The fast rotation effects are taken into account with \omc=95\%.
The typical errors are shown in the lower left corner. The diagonals {\revised show the area of
existing Be stars in the MW (Zorec et al. 2005).} 
}
\label{statutevolBeSMC}
\end{figure}

To investigate the evolutionary status of Be stars in the SMC taking into account the effects of fast
rotation and the low metallicity (Z=0.001) of the SMC, we firstly calculate the lifetime of the main sequence
(\tms) of massive stars  by interpolation in grids of evolutionary tracks provided by Maeder \& Meynet (2001) and
Meynet \& Maeder (2000) for  different metallicities, with an initial velocity V$_0$ = 300 \kms~and for stars
with masses higher than 9 M$_{\odot}$.  Secondly, we  extrapolate the \tms~values towards lower masses (5-9
M$_{\odot}$). We then investigate the evolutionary status \ttms~of the SMC Be stars  using the values of their
ages corrected from fast rotation effects and given in Table~\ref{LMRBePNM95}. Results are shown in
Fig.~\ref{statutevolBeSMC}.

The following remarks can be made:
\begin{itemize}
\item It appears that
more massive Be stars in the SMC are
evolved: all of them in our sample are localized in the second part of
the MS.
\item Intermediate mass Be stars 
are scattered across the MS. 
\item Less massive Be stars are mainly evolved and in the second part 
of the MS (\ttms$\ge$ 0.5).
\end{itemize}

\begin{figure}[!htbp]
\centering
\resizebox{\hsize}{!}{\includegraphics[angle=-90]{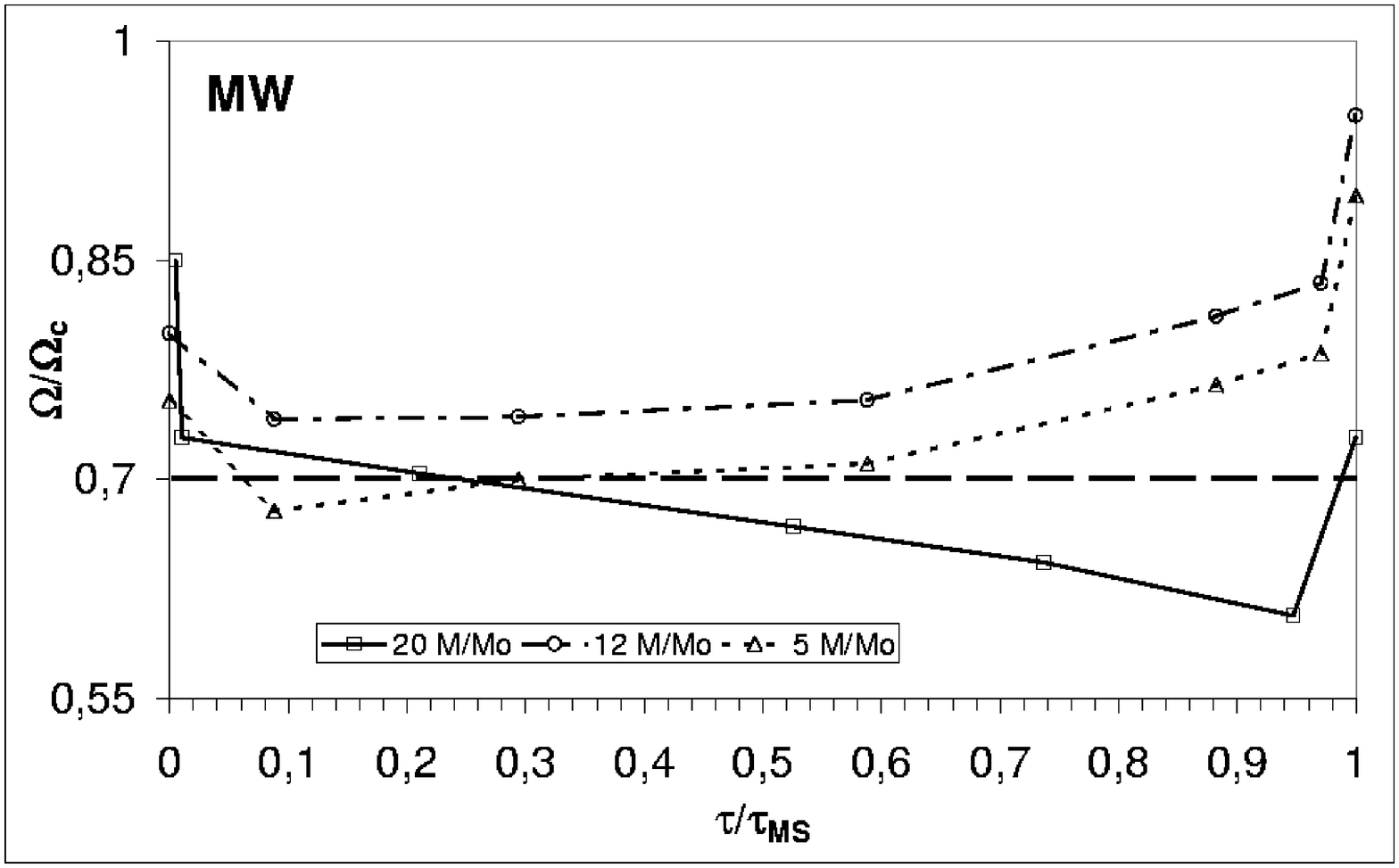}}\\%MWomegaBe.ps}}
\resizebox{\hsize}{!}{\includegraphics[angle=-90]{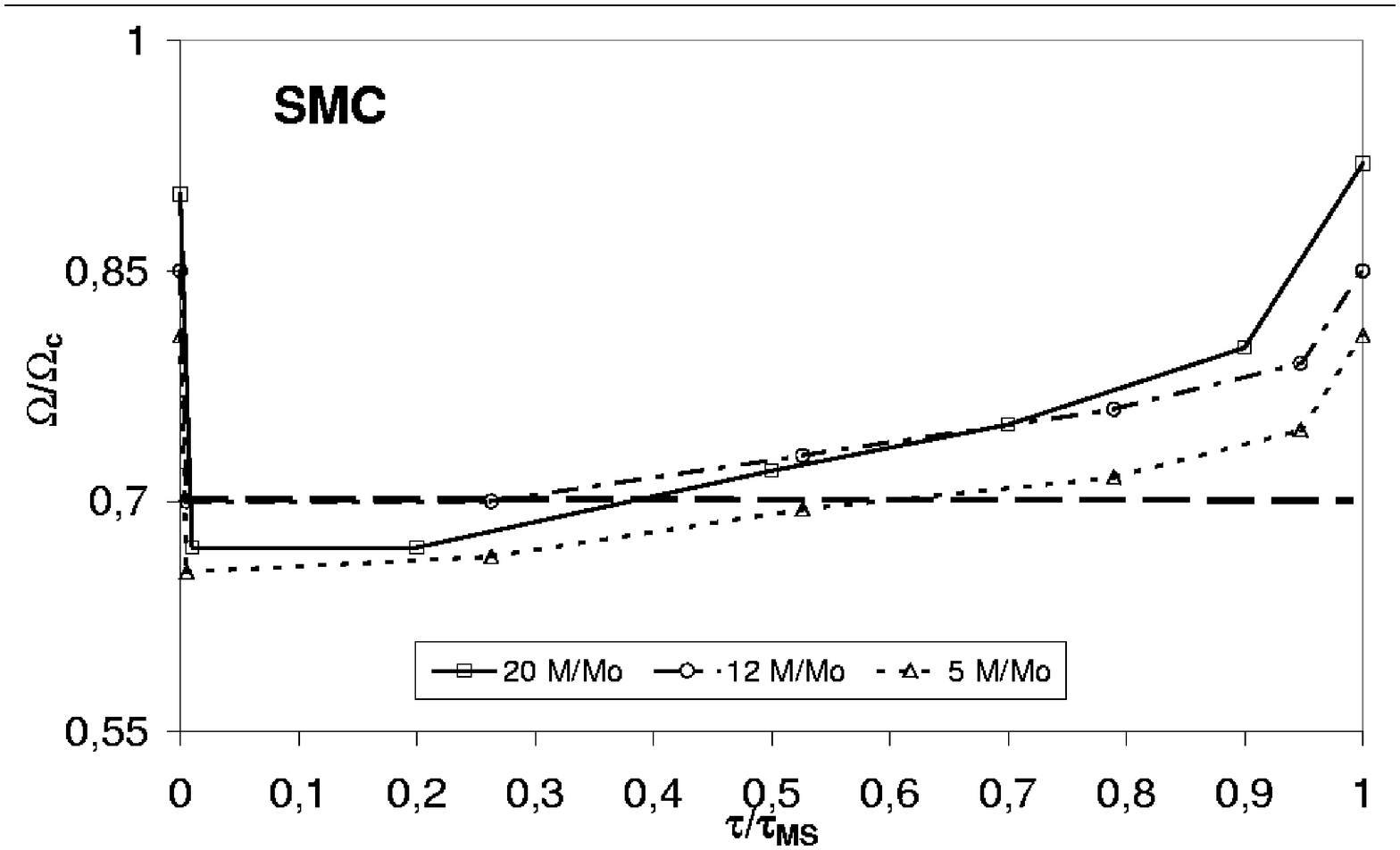}}%SMComegaBe.ps}}

\caption{Evolution of the \omc~ratio for 3 types of stars:
20 M$_{\odot}$ (squares), 12 M$_{\odot}$ (circles) and 5 M$_{\odot}$ (triangles) 
stars in the MW (top) and in the SMC (bottom).
The stars may become Be stars if \omc$\geq$70\% (noted with a long dashed line).}
\label{evolBeOMmoa}
\end{figure}
\begin{table}[th]
\caption{Proportions in SMC, LMC, and MW of Be stars (in \%) in the upper (\ttms$>$ 0.5) and lower (\ttms$\leq$ 0.5) MS 
for masses $>$ 12 M$_{\odot}$ and $\leq$ 12 M$_{\odot}$. The values for the MW are taken from Zorec et al. (2005). }
\centering
\begin{tabular}{@{\ }c@{\ \ \ }c@{\ \ \ }c@{\ \ \ }c@{\ \ \ }c@{\ \ \ }c@{\ \ \ }c@{\ \ \ }c@{\ \ \ }c@{\ \ \ }c@{\ }}
\hline
\hline
 & M $>$ 12 M$_{\odot}$ & \\
\hline
 & MW & LMC & SMC \\
\ttms$>$0.5 & 30 & 100 & 100 \\
\ttms$\leq$0.5 & 70 & 0 & 0 \\
\hline
 & M $\leq$ 12 M$_{\odot}$ & \\
\hline
 & MW & LMC & SMC \\
\ttms$>$0.5 & 65 & 77 & 94 \\
\ttms$\leq$0.5 & 35 & 23 & 6 \\
\hline
\end{tabular}
\label{propBe}
\end{table}

\subsection{Comparison between the evolutionary status of Be stars in the MC and MW}

To compare the evolutionary status of Be stars in the different galaxies we calculate the \ttms~ratio for massive
and less massive stars  in the LMC (Z=0.004) and the MW (solar metallicity) as for the SMC. We note that in the
MW, our results are quite similar to the ones in Zorec et al. (2005). However, in the LMC, our previous results
shown in PaperI (Fig.12) are slightly modified with the use of evolutionary tracks  adapted to its metallicity
(Z=0.004), mainly for the less massive Be stars which appear more evolved in the present study.  The proportions
of more massive ($>$12 M$_{\odot}$) and less massive ($\leq$12 M$_{\odot}$) Be stars in the SMC, LMC and MW are
summarized in Table~\ref{propBe}. It is shown that all more massive stars in the LMC and the SMC are in the upper
part of the MS (\ttms$\ge$ 0.5), contrary to the MW where they are mainly in the lower part of the MS. The less
massive stars seem to follow the same trend in the MW and the MC, they are mainly in the upper part of the MS in
agreement with Fabregat \& Torrejon (2000). With our results on the ZAMS rotational velocity distributions
(Fig.~\ref{V0ZAMSequawind}) and with the theoretical evolution of angular velocities taken from the Geneva models
mentioned above, we obtain the evolution of the angular velocities for different masses (5, 12, 20 M$_{\odot}$)
of Be stars in the MW and the SMC as it is shown in Fig.~\ref{evolBeOMmoa}. We have emphasized significant
behaviour differences between Be stars in function of their mass and of their metallicity environment, and we
propose the following explanation of the Be phenomenon  in the MW and in the MC:

\begin{itemize}
\item More massive Be stars:\\
In the MW, more massive stars begin their life in the MS with a
high \omc~($>$70\%, see Fig.~\ref{V0ZAMSequawind} and Table~\ref{omegaZAMSBe}), 
thus these stars could be Be stars. Then, by angular momentum loss, the stars spin 
down and might not eject matter anymore; thus, they lose their ``Be star'' character during the
first part of the MS (typically for \ttms$>$ 0.2). However, at the end of
the MS (\ttms$\simeq$ 1), during the secondary contraction, massive stars could
have \omc~sufficiently high to become Be stars again.

In the MC, more massive stars begin their life in the MS with a high \omc~($>$70\%)
and could be Be stars but only at the very beginning of the ZAMS. Therefore, they rapidly 
lose ``the Be star status''. However, they spin
up (increase of the radius, low mass-loss and low angular-momentum loss) and
reach very high \omc~at the end of the  first part of the MS
(typically after \ttms$>$ 0.4) so that they become Be stars again, contrary
to what is  observed  in the MW. 
\item Intermediate mass Be stars:\\
In the MW as in the MC,  intermediate mass Be stars  begin their life in the MS
with high \omc~($>$70\%, Fig.~\ref{V0ZAMSequawind} and Table~\ref{omegaZAMSBe}), therefore the stars
could be Be stars. Then, the evolution of their \omc~allows these
stars to remain Be stars.
\item Less massive Be stars:\\
In the MW as in the MC, less massive stars begin their life on the MS with
\omc~sufficiently high to obtain the status of ``Be stars''. After the
fast internal angular redistribution in the first $\simeq$10$^{4}$ years in the ZAMS,
the surface rotational velocities decrease and the stars lose the status of ``Be
stars''. Then, the evolution of their \omc~allows these stars to
become Be stars again (typically after \ttms$>$ 0.5), in agreement with
Fabregat \& Torrej\'on (2000). 
\end{itemize}

We note that our propositions explain results presented by Zorec et al. (2005, their Fig. 6) and our
results concerning the LMC (Paper I) and the SMC (Fig.~\ref{statutevolBeSMC}). We also remark no difference
between the evolutionary status of Be stars in clusters and fields.

In summary, a B star can become a Be star if its progenitor has a strong
initial rotational velocity at the ZAMS. Depending on the metallicity of the
environment of a star and on the stellar mass, the Be phenomenon appears at
different stages of the MS.

%@@@@@@@@@@@@@@@@@@@@@@@@@@@@@@@@@@@@@@@@@@@@@@@@@@@@@@@@@@@@@@@@@@@@@@@@@@@@@@

%@@@@@@@@@@@@@@@@@@@@@@@@@@@@@@@@@@@@@@@@@@@@@@@@@@@@@@@@@@@@@@@@@@@@@@@@@@@

\section{Conclusions}

With the VLT-GIRAFFE spectrograph, we obtained spectra of a large sample of B
and Be stars in the SMC-NGC\,330 and surrounding 
fields. We determined fundamental parameters for B stars in the sample, 
and apparent and parent non-rotating counterpart (pnrc) fundamental parameters 
for Be stars.

Using results from this study and
those obtained for the LMC with the same instrumentation (Paper I), we made a
statistical comparison of the behaviour of B and Be stars in the Magellanic
Clouds with the MW.\\
\begin{itemize}
\item there is no difference in rotational velocities between early-type stars in clusters 
and in fields in the LMC and in the SMC.
\item the lower the metallicity is, the higher the rotational velocities are.
B and Be stars rotate faster in the SMC than in the LMC, and faster in the LMC
than in the MW. 
\item we have determined, for the first time, the distributions of the ZAMS 
rotational velocities of Be stars. The ZAMS rotational velocities are mass-dependent
and metallicity-dependent. The gradients of the distributions seem to be similar 
whatever the metallicity.
\item Only a fraction of B stars, that reach the ZAMS with sufficiently high initial
rotational velocities, can become Be stars. Wisniewski \& Bjorkman (2006) seem to have found
photometrically a similar result.
\item the angular velocities are similar for B stars in the LMC and in the 
MW and lower than those in the SMC. The same result is obtained for Be stars.
\item more massive Be stars in the SMC, which are evolved, are critical rotators.
\item in an evolutionary scheme, massive Be stars in the MC and more particularly in the SMC
appear in the second part of the main sequence, to the contrary to massive Be stars in the MW,
which appear in the  first part of the main sequence.
\item other categories of Be stars (intermediate and less massive Be stars)
follow the same evolution whatever the metallicity of the environment.
\end{itemize}

Our results support Stepi\'en's scenario (2002): massive stars with a weak or
moderate magnetic field and with an accretion disk during at least 10\% of
their PMS lifetime would reach the ZAMS with sufficiently high initial
rotational velocity to become Be stars.

Our observational results also support theoretical results by Meynet \& Maeder
(2000, 2002) and Maeder \& Meynet (2001) for massive stars. In a more general way
they illustrate the behaviour of massive stars and the importance of the processes
linked to rotation and to metallicity.  In particular, our results show that
rotation has a major role in the behaviour of massive stars. Moreover, this
pioneer study of a large sample of B stars in low metallicity environments, such
as the LMC (Paper I) and the SMC (this study), allows us to approach  the ``first
stars''. We can reasonably expect that first massive stars are fast
rotators.

In forthcoming papers, we will present results of CNO abundances determinations, discuss the proportion of Be
stars and report on the discovery of binaries in the SMC.

%@@@@@@@@@@@@@@@@@@@@@@@@@@@@@@@@@@@@@@@@@@@@@@@@@@@@@@@@@@@@@@@@@@@@@@@@@@
\footnotesize{
\begin{acknowledgements}
We would like to thank Dr V. Hill for performing the observing run in
September 2004 with success and good quality. We thank Drs C. Ledoux, P. Fran\c{c}ois, and E.
Depagne for their help during the observing run in October 2003.
This research has made use of the Simbad database and Vizier
database maintained at CDS, Strasbourg, France, as well as of the WEBDA
database.
\end{acknowledgements}

}

\end{document}